\documentclass[final,3p,times]{elsarticle}
\biboptions{sort&compress}

\usepackage{hyperref}

\usepackage{graphicx}

\usepackage{amssymb}
\usepackage{amsmath}
\usepackage{upgreek}
\usepackage{lineno}
\usepackage{gensymb}
\usepackage[utf8]{inputenc}
\usepackage[dvipsnames]{xcolor}

\usepackage{fancyhdr}
\usepackage[normalem]{ulem}

    \makeatletter
    \def\ps@pprintTitle{%
       \let\@oddhead\@empty
       \let\@evenhead\@empty
       \let\@oddfoot\@empty
       \let\@evenfoot\@oddfoot
    }
    \makeatother

\journal{Ultramicroscopy}

\begin{document}

\Large
 \color{red}
This is the Preprint version of our article accepted for publication in Ultramicroscopy \citep{Hettler.2021}. The Version of Record is available online: \url{https://doi.org/10.1016/j.ultramic.2021.113319}.
\color{black}
\normalsize
\begin{frontmatter}

\title{Comparative image simulations for phase-plate transmission electron microscopy}

\author[label1]{Simon Hettler\corref{cor1}}
\author[label1,label2,label3]{Raul Arenal}
\address[label1]{Laboratorio de Microscopías Avanzadas (LMA), Universidad de Zaragoza, Zaragoza, Spain}
\address[label2]{Instituto de Nanociencia y Materiales de Aragón (INMA), CSIC-Universidad de Zaragoza, Zaragoza, Spain}
\address[label3]{ARAID Foundation, Zaragoza, Spain}
\cortext[cor1]{hettler@unizar.es}

\begin{abstract}
    Numerous physical phase plates (PP) for phase-contrast enhancement in transmission electron microscopy (TEM) have been proposed and studied with the hole-free or Volta PP having a high impact and interest in recent years. This study is concerned with comparative TEM image simulations considering realistic descriptions of various PP approaches and samples from three different fields of application covering a large range of object sizes. The simulated images provide an illustrative characterization of the typical image appearance and common artifacts of the different PPs and the influence of simulation parameters especially important for PP simulations. A quantitative contrast analysis shows the superior phase-shifting properties of the  hole-free phase plate for biological applications and the benefits of adjustable phase plates. The application of PPs in high-resolution TEM imaging, especially of weak-phase objects such as (atomically thin) 2D materials, is shown to increase image interpretability. The software with graphical user interface written and used for the presented simulations is available for free usage.
\end{abstract}

\begin{keyword}
transmission electron microscopy \sep phase plate \sep phase contrast \sep image simulation \sep software
\end{keyword}
\end{frontmatter}

\section{Introduction}

Physical phase plates (PPs) for phase-contrast enhancement are well known since the beginnings of transmission electron microscopy (TEM) \citep{Boersch_PP_1947}. PPs induce a relative phase shift between the scattered and unscattered parts of the electron wave and thus offer the possibility to make phase information of so-called weak-phase objects (WPOs) visible without the need to defocus. The phase shift is typically applied in the back focal plane (BFP) of the objective lens, where the zero-order beam (ZOB = the unscattered part) and scattered electrons are spatially separated. While ideal PPs provide a phase shift that is either only affecting the ZOB (located at spatial frequency $u$ = 0) or the scattered parts ($u>$ 0), all experimental PP designs proposed so far induce a phase shift that is spatially distributed over both parts. The reason is the small dimension of the ZOB of typically below 100~nm and the challenging task to fabricate a device that generates a suitable highly localized phase-shifting field. 

Different PPs have been proposed and realized experimentally \citep{PPReviewMalacHettler} among which the Volta or hole-free (HF)PP so far showed the highest impact due to a very narrow phase-shift distribution formed on a thin film by the ZOB itself \citep{HFPP_2012,VPP_2014}. The PP closest to its light optical analog \citep{ZernikeLight} is the Zernike PP, where the scattered electrons experience a phase shift due to the mean inner potential of an amorphous carbon thin film with a central hole \citep{NagayamaDanev2001b}. Further PPs are based on microfabricated devices that create electrostatic potentials or currents close to the ZOB, such as the Boersch \citep{BoerschOptimize.2007} and Zach PP \citep{PPZach_MM2010} or the Tunable Ampere (TA)PP \citep{Marco_TunablePP_2018}. Only recently, a PP based on the interaction between an intense laser field and the electron beam has been demonstrated experimentally \citep{Glaeser_LaserPP_2019}.

Due to the presence of obstructing or damping elements, electrostatic charging and/or the spatial phase shift distribution of the PP design differing from the ideal case, all PPs introduce image artifacts. These artifacts strongly depend on the PP type and as well on the investigated object and can either dominate the image contrast as it often is the case for, e.g., Boersch PPs \cite{ZPP_Glaeser_Boersch_UM_2007,ZPP_Boersch_UM_2010}, or be almost invisible, manifesting itself only in a halo around object borders for HFPPs \citep{Pretzsch.2019}. HFPPs so far seem to offer the best phase-shifting properties and have been applied in imaging of WPOs, such as biological \citep{PPFukudaCell_2015,MatthiasEbolaNature2018} or magnetic samples \citep{2019_Kotani_SkyrmionHFPP}. Electrostatic PPs have also been employed in high-resolution (HR)TEM \citep{ZPP_Boersch_UM_2010,SimonZachPP_2016} and may improve imaging in combination with aberration correction \citep{Gamm.2008EffectPP} and provide the possibility of an object-wave reconstruction \citep{PPZach_MM2010,Gamm.2010}. 

In order to get to a qualitative or semi-quantitative interpretation of the acquired phase-contrast images, image simulations have frequently been conducted to disentangle the introduced artifacts from the actual phase contrast \citep{ZPP_Boersch_UM_2010,Dries.2014,Marco_TunablePP_2018,Pretzsch.2019,Glaeser_LaserPP_2019,Obermair2020,HaradaGrating_2020}. Especially the Zernike PP has been subject of several theoretical studies dealing with the expectable contrast and optimum parameters \citep{BeleggiaZernikeFormula2008,Fukuda.2009,ZPPOptimizing_NagayamaDanev_UM2011,ZPPEdgcombe_2014}. The present study deals with comparative TEM image simulations of three test objects with varying feature size, which cover different fields of application where the PP is or could be beneficially applied. The simulations are conducted using realistic representations of several PP designs, which have already been demonstrated experimentally. The results show, in an illustrative way, the typical image appearance for the different PP designs. A quantitative contrast and visibility analysis allows the comparison of these different PP approaches. The importance and impact of several simulation parameters are assessed. Especially the consideration of a finite diameter of the ZOB is essential to obtain a better match with experiments. The program written for the conducted simulations, described in the supplementary information (section \ref{SIS9}), is made available for free \citep{PPSim2}.

\section{Simulation of phase-plate TEM images}
\label{S:Theory}

\subsection{Selected specimen wave functions}
\label{S:WaveFun}
Starting point of the image formation is the complex object exit-wave function (OEWF):

\begin{equation}
\Psi_O(x,y)=a(x,y) \exp(i\xi (x,y))    
\end{equation}

given by the object-induced phase $\xi$ and amplitude $a$  of the electron wave in the two-dimensional plane after transmission through the specimen given by spatial cartesian coordinates ($x$, $y$). OEWFs can be determined via multi-slice simulations \citep{Rosenauer.2008STEMSim,STEMCELLGrillo.2013,Barthel.2018}. For amorphous or biological samples, such as ice-embedded cryo samples or amorphous C films, the approach of a mean inner potential (MIP) $V_{MIP}$ is commonly used. In this approach, the phase shift $\xi (t)$ induced by a material with a local thickness $t$ is given by:

\begin{equation}
    \xi (t(x,y)) = \sigma \cdot t(x,y) \cdot V_{MIP} = \frac{\pi e 2 (E_0+E)}{\lambda E (2E_0+E)} \cdot t(x,y) \cdot V_{MIP}
    \label{Eq:PhaseMIP}
\end{equation}

with the interaction constant $\sigma$ depending on the electron rest energy $E_0$, its kinetic energy $E$ and the wavelength $\lambda$. Assuming an energy-filtered image in which all inelastically scattered electrons are taken out by an energy filter, the amplitude is estimated using an inelastic mean free path ($\lambda_{MFP}$):

\begin{equation}
    a (t(x,y)) = \exp{(-t(x,y)/\lambda_{MFP}})
    \label{Eq:AmpMFP}
\end{equation}

The following three test specimens were used to evaluate the PP performance over a range of different object sizes:

\begin{enumerate}
    \item Two T4 bacteriophages (approximated as cylindrical structure ($V_{MIP}$ = 6~V) with a diameter of 85~nm) embedded in 150~nm of vitrified ice ($V_{MIP}$ = 3.5~V \citep{harscher1998inelasticMIP}) as described in \citep{Obermair2020}. In this OEWF, the phages and the vitrified ice are treated as pure phase objects. The T4 bacteriophages are an example for a macromolecular complex in biological applications.
    \item Spherical amorphous carbon (aC) nanoparticles (NPs) ($V_{MIP}$ = 9~V) with sizes between 1 and 20 nm embedded in 30~nm of vitrified ice ($V_{MIP}$ = 3.5~V). Ice and NPs are assumed to have the same $\lambda_{MFP}$~=~300~nm. This OEWF is thought as an application example for single-particle analysis and, due to their homogeneous shape, the NPs serve to visualize PP-induced effects.
    \item A monolayer of graphene obtained by multislice calculation with STEMsim \citep{Rosenauer.2008STEMSim}, as an example for a nano science application.
\end{enumerate}

\begin{figure}
\centering
    \includegraphics[width=\linewidth]{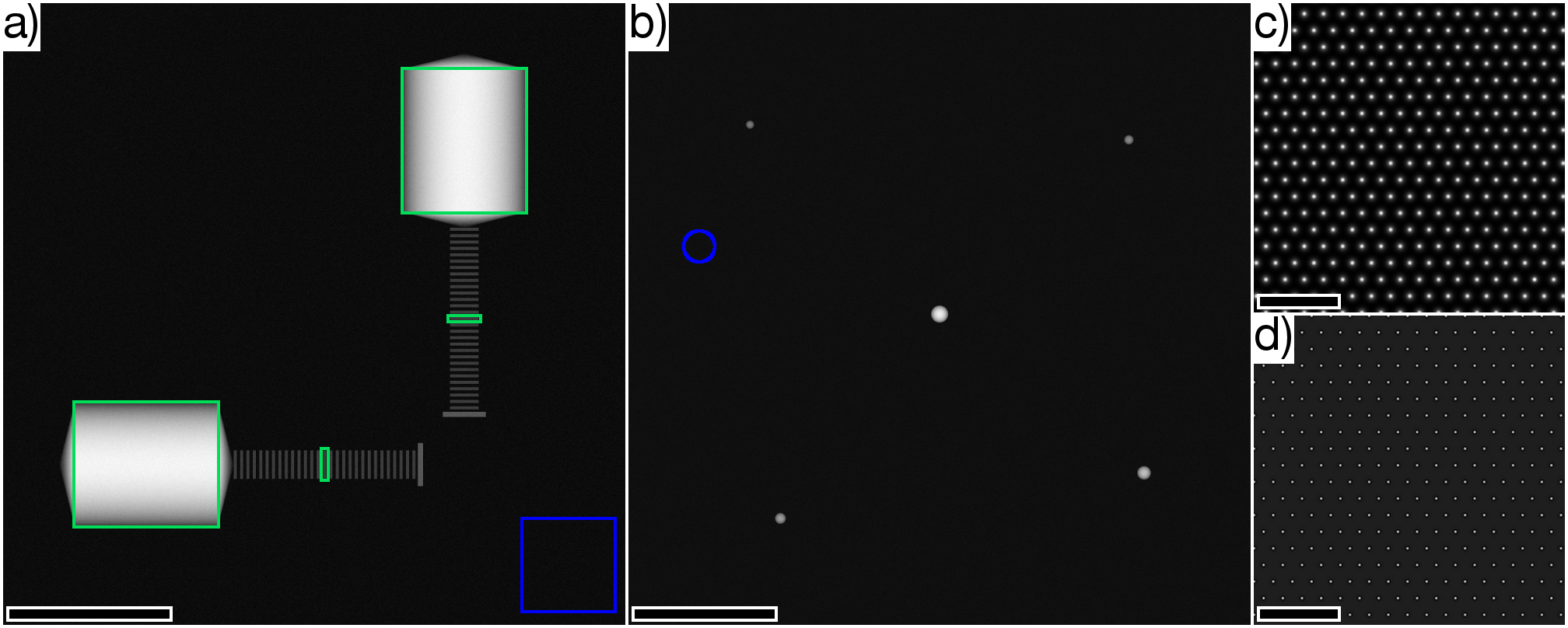}
    \caption{OEWFs used for image simulations. (a) Phase of two T4 bacteriophages \citep{Obermair2020}. (b)  Phase of five spherical amorphous C NPs with diameters of 6, 7, 8, 10 and 12.5~nm embedded in 30~nm of vitrified ice. (c) Phase and (d) amplitude of monolayer graphene. (c,d) show a part of the entire OEWF for sake of visibility. Scale bars are (a,b) 100~nm and (c,d) 1~nm. Contrast scales from black to white are (a) 3.62 [1.15] - 4.99 [1.59]~rad [$\uppi$], (b) 0.67 [0.19] - 0.91 [0.29]~rad[$\uppi$], (c) 0 - 0.22 [0.07]~rad[$\uppi$] and (d) 0.99 - 1.10. Green and blue marked areas correspond to object and background areas selected for contrast and visibility quantification. (For interpretation of the references to colour in this figure legend, the reader is referred to the web version of this article.)}
    \label{F:OEWFs}
\end{figure}

The phase of the T4 bacteriophages is depicted in Figure~\ref{F:OEWFs}a, where a noise of 3.5\% was added to the OEWF phase. The aC NPs are distributed over three OEWFs to avoid overlap of fringing contrast of the different NPs in the simulated images. Figure~\ref{F:OEWFs}b shows the phase of the OEWF with the NP diameters between 6 and 12.5~nm. A noise of 4\% and 2\% was added to phase and amplitude, respectively, to reflect the amorphous structure of the vitrified ice. Phase and amplitude of a part of the graphene monolayer OEWF are shown in Figure~\ref{F:OEWFs}c, d. In this case, no noise has been added to the OEWF.

\subsection{Image formation}
\label{S:ImgFor}

Theory of linear image formation may be found in textbooks \citep{ReimerTEM_5,RosenauerBook.2003} and this Section only presents the formulas used for the simulations carried out in this work. Objective and projector lenses form an image wave $\Psi_I(x,y)$ in the plane of the detector. The aberrations of the objective lens, possibly modified by an aberration corrector, which are imposed on the phase of the electron wave, are described by the wave aberration function $\chi (\vec{u})$ in dependence of the spatial frequency $\vec{u} = (u_x, u_y)$. In the presence of a physical phase plate, $\chi (\vec{u})$ is given by:

\begin{equation}
    \chi(\vec{u}) = \pi Z \lambda u^2 + \frac{\pi}{2} C_S \lambda^3 u^4 + \phi_{PP} (\vec{u}) + ...
    \label{Eq:WaveAb}
\end{equation}

with the defocus $Z$, the electron wavelength $\lambda$, the spherical aberration constant $C_S$ and the phase shift introduced by the PP $\phi_{PP} (\vec{u})$, discussed in Section~\ref{S:PPs}. Higher order aberrations are neglected for the phages and aC NPs and are set to the values as detailed in the SI (section \ref{SIS1}) for the graphene sample to reflect common values for an aberration-corrected microscope.

$\Psi_I(x,y)$ is calculated by the weighted-focal series to account for partial temporal coherence. In this approach, $\Psi_I(x,y)$ is an incoherent weighted sum of images calculated with varying defocus $Z + \delta Z$. In the presented simulations, we calculated 11 images for varying $\delta Z$, which are distributed over a range of $3\Delta$, with the focal spread $\Delta$ depending on the variations of accelerating voltage, electron energy and lens currents. The intensity $I(x,y)$~=~$|\Psi_I(x,y)|^2$ measurable by a suitable detector, e.g. by a camera or a fluorescence screen, is given by:

\begin{equation}
    I(x,y) = \sum_{\delta Z} \frac{B}{2 \pi \Delta } exp\left[-\frac{\delta Z^2}{2 \Delta ^2}\right] \cdot |  \mathcal{F}^{-1} \left\{ \mathcal{F} \{\Psi_O(x,y)\} \cdot exp\left[-i\chi (\vec{u},Z+\delta Z)\right] \cdot E_S(\chi(\vec{u},Z+\delta Z)) \cdot A_{PP}(u) \right\}|^2
    \label{Eq:ImageFormation}
\end{equation}

with $\mathcal{F}$ ($\mathcal{F}^{-1}$) the (inverse) Fourier transformation and $A_{PP}(u)$ the aperture function being 1 below and 0 above the radius $u_{OA}$ of the objective aperture and considering possible damping or obstructing structures of the PPs (see Section~\ref{S:PPs}). $B$ is the normalization factor for the weighted focal series determined by $\sum_{\delta Z} B / (2 \pi \Delta ) \cdot exp\left[-\delta Z^2 /(2 \Delta ^2)\right] = 1$.  Partial spatial coherence for small semi-convergence angles $\alpha$ ($<$0.5 mrad) may be considered by multiplication in reciprocal space with the envelope $E_S$:

\begin{equation}
    E_S (\vec{u})= exp \left( - \frac{\alpha^2}{4 \lambda^2} \left(\frac{\partial \chi (\vec{u})}{\partial \vec{u}}\right)^2 \right)
    \label{Eq:EnvelSpatialCoherence}
\end{equation}

Upon detection of the electron wave by a camera, the electron wave is transformed in a pixelated image by counting the electrons hitting the detector. Experimental Poisson noise may be considered by using the simulated pixel size and the applied electron dose in electrons per unit area. Upon detection, the electron wave is further modified by the modulation transfer function (MTF) \citep{Thust.2009}. Experimental noise and MTF are neglected in the present study as they do not alter the comparison between different PPs. In case of the T4 and the spherical aC NPs, the effect of noise on the image appearance is shown in the SI for exemplary images (Fig. \ref{SF:T4_Noise} and \ref{SF:Sp_Noise}).

\subsection{Consideration of a physical phase plate}
\label{S:PPs}

The phase profile of the PPs is considered by adding $\phi_{PP} (\vec{u})$ to the wave aberration function (Eq.~\ref{Eq:WaveAb}) and damping or obstruction are taken into account by a multiplicative factor in $A_{PP}(u)$. To obtain these phase and amplitude profiles, they first are calculated in real space and are then transformed to the reciprocal space depending on the electron wavelength $\lambda$ and the focal length of the BFP in which the PP is positioned (see Section \ref{S:Mics}).

The PP profiles are implemented following their physical appearance, i.e. they are not normalized with respect to the phase shift induced on the ZOB. As thus the term positive or negative phase shift depends on the used PP and may lead to confusion, the expression phase contrast (PC) is referred to when discussing the results. Dark or bright appearance against the background is referred to as positive or negative PC, respectively \citep{HFPP_Pol2017}. Positive PC is achieved by inducing a negative phase shift on the ZOB with respect to the scattered electrons and vice versa.

In conventional (C)TEM, the Scherzer defocus $Z_S$ is used to obtain a broad band of strong phase contrast transfer. In the presence of an additional phase shift of $\uppi$/2 induced by an ideal PP, this transfer band shifts and an optimum transfer is reached at a smaller $Z_{S,PP}$~=~-0.73$\sqrt{\lambda C_S}$ for an applied phase shift of $\uppi$/2 \citep{NagayamaDanev2001b}. This PP Scherzer focus was used for the simulations of the PP images.

\subsubsection{Studied phase-plate designs}

\begin{figure}
    \centering
    \includegraphics[width=0.5\linewidth]{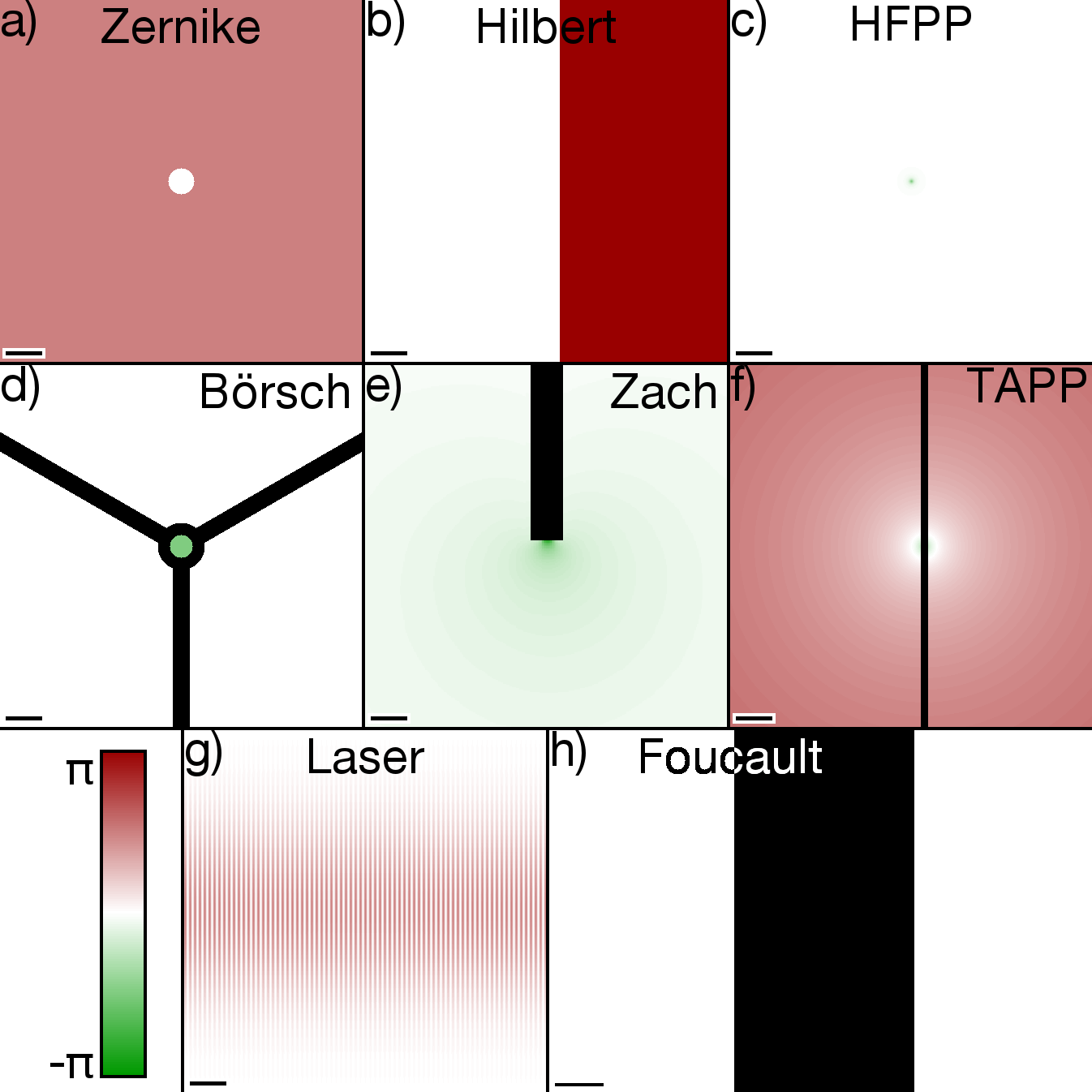}
    \caption{Studied phase-plate designs including thin-film based (a) Zernike, (b) Hilbert and (c) hole-free PPs; electrostatic (d) Boersch and (e) Zach PPs at -$\uppi$/2; (f) the tunable Ampere PP ($I_{TAPP}$~=~2.5~mA); (g) the laser PP at $\uppi$/2 and (h) the blocking Foucault aperture. For this figure, the PP profiles have been calculated using the \textit{cryo-Mic} (Table~\ref{T:Mics}), with focal lengths of (a-c,f,h) 5~mm, (d,e) 15.8~mm and (g) 20~mm. The scale bars correspond to a spatial frequency of (a-g) 0.1~nm\textsuperscript{-1} and (h) 0.4~nm\textsuperscript{-1}. Obstructing elements are displayed in black, the phase in a scale between -$\uppi$ (green), 0 (white) and $\uppi$ (red). (For interpretation of the references to colour in this figure legend, the reader is referred to the web version of this article.)}
    \label{F:PPs}
\end{figure}

\begin{figure}
    \centering
    \includegraphics[width=0.9\linewidth]{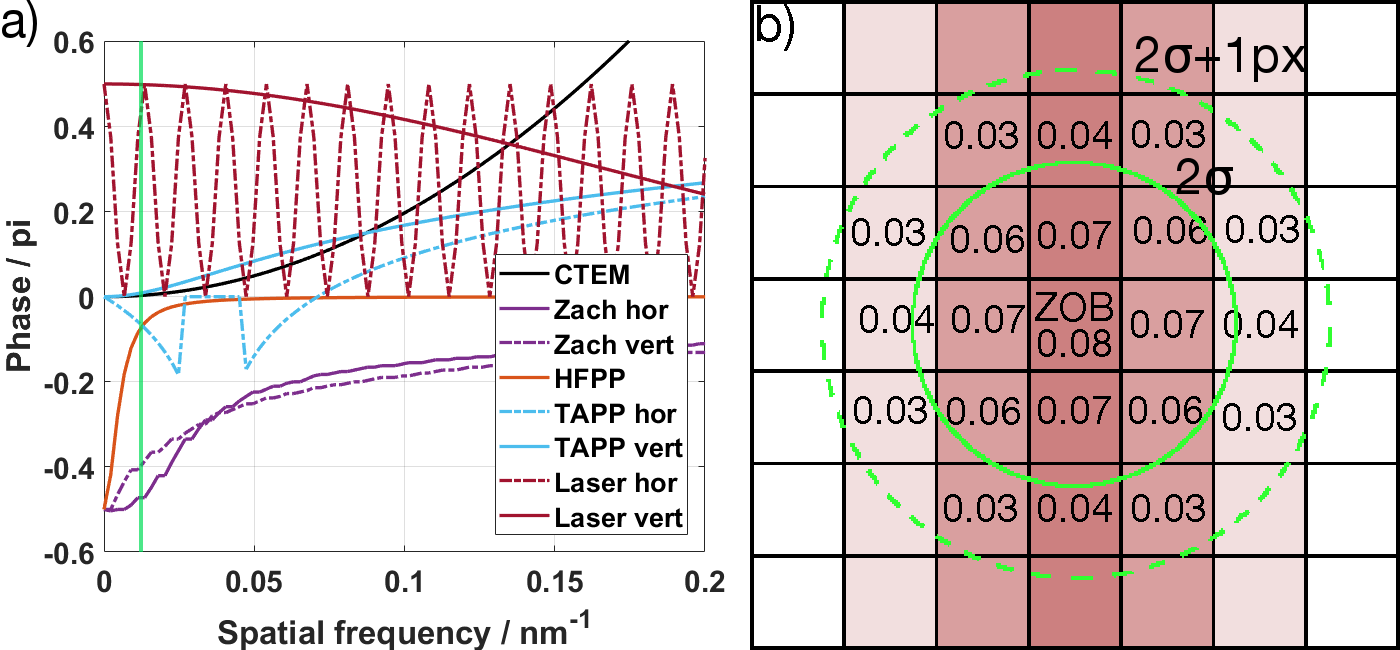}
    \caption{(a) Comparative plot of phase-shifting profiles of PPs with a smooth or varying profile including conventional (C)TEM at a defocus of -10~$\upmu$m (black solid line), the HFPP (orange line) and the horizontal (solid lines) and vertical (dash-dotted lines) profiles of the Zach PP (purple), the TAPP for $I_{TAPP}$~=~2.5~mA (light blue) and the LPP (red) calculated for the simulations of phages and aC NPs. The green vertical bar marks the characteristic spatial frequency of the phage body. (b) Schematic visualization of the consideration of a finite ZOB in pixel-based simulations. The scheme represents the central 7x7 pixels of the LPP phase shift from Figure~\ref{F:PPs}g and a ZOB with an exemplary $\sigma$~=~40~nm (0.08 mrad) in the BFP (=~160~nm in the laser BFP) marked by the solid green ring. All pixels being by more than 50~\% within the dashed ring are considered for the simulations of the T4 phages. The values represent the weighting factor for the image simulated with the corresponding pixel used as ZOB. (For interpretation of the references to colour in this figure legend, the reader is referred to the web version of this article.)}
    \label{F:PPLineScans}
\end{figure}

This study is carried out using eight different PPs with real-space profiles defined by either cartesian ($\hat{x}$,$\hat{y}$) or polar coordinates ($\hat{r}$,$\hat{\theta}$) as detailed below and displayed in Figure~\ref{F:PPs} and \ref{F:PPLineScans}a. Without claiming completeness, only PPs, which have been realized experimentally, have been considered. This includes three thin-film based PPs, three electrostatic or current-carrying PPs, the laser PP and an obstructing aperture. The PPs were considered in their standard implementation and for possible design optimizations and other approaches we refer to the literature \citep{PPReviewMalacHettler}. All PPs based on thin films are assumed to be made of aC with $V_{MIP}$~=~9~V and $\lambda_{MFP}$~=~180~nm. $V_{MIP}$ is used to determine the required thickness for the thin-film based PPs (Eq.~\ref{Eq:PhaseMIP}) and the corresponding amplitude damping is calculated by Eq.~\ref{Eq:AmpMFP}.

The lower value of $\lambda_{MFP}$ in comparison with the one used for the OEWF containing aC NPs is chosen due to the fact that in addition to inelastic scattering, also elastic scattering in the PP leads to an effective reduction in intensity. The reason is that electrons scattered (or diffracted) in the PP film deviate from their original path leading to a shift in the image plane, an effect which has been studied for crystalline PPs \citep{Dries.2014}. Characteristic elastic scattering angles are relatively large (68~mrad for C at 300 kV, see chapter 3.1.2 in \citep{Egerton.2011}) and this leads to a strong shift in the image plane  (340~$\upmu$m for 68~mrad with a focal length of 5~mm). Under the assumption that a sample area not significantly larger than the final image on the camera is illuminated, only a negligible amount of electrons scattered elastically in the PP will end up on the camera justifying the assumption of a complete removal of any scattered electron in the PP film.

\paragraph{Thin-film based Zernike PP}

The Zernike (Z)PP consists of a thin film (thickness $t_{ZPP}$) of aC with a central hole with a radius $r_{ZPP}$ (Figure~\ref{F:PPs}a) \citep{NagayamaDanev2001b}. Different values were considered for $r_{ZPP}$ but the presented results in the main manuscript represent data calculated with $r_{ZPP}$~=~350~nm. Although smaller values of $r_{ZPP}$~=~250~nm have been used experimentally \citep{Fukuda.2009}, the chosen value is a compromise between a good cut-on frequency and alignment requirements. The phase profile is given by $\phi_{ZPP} (\hat{r}\leq r_{ZPP}$)~=~0 and $\phi_{ZPP} (\hat{r}>r_{ZPP})$~=~$\pi$/2. Zernike PP images show positive PC.

\paragraph{Thin-film based Hilbert PP}

The Hilbert (H)PP \citep{Nagayama_JBP_2002} as well consists of a thin film of aC with a thickness $t_{HPP}$, which covers only half of the plane (Figure~\ref{F:PPs}b). The PP is positioned with a fixed distance $d_{HPP}$ (350~nm) between ZOB and the film onset. The phase profile is given by $\phi_{HPP}(\hat{x} \leq d_{HPP})$~=~0 and $\phi_{HPP} (\hat{x}>d_{HPP})$~=~$\pi$.

\paragraph{Thin-film based hole-free or Volta PP}

The hole-free (HF)PP consists of a continuous aC thin film on which the impinging ZOB creates a phase-shifting profile \citep{HFPP_2012,VPP_2014}. The most probable explanation is an effective change of the aC work function in the area irradiated by the ZOB caused by the electron-stimulated desorption of a chemisorbed water layer \citep{2018_Simon_NegCharge,2019_Simon_LN2_Charging}. This leads to a patch with negative potential and the resulting phase profile $\phi_{HFPP}$ may be approximately described by a Lorentzian distribution:

\begin{equation}
\phi_{HFPP} (\hat{r}) =  \frac{\varphi_{HFPP}}{(2\cdot \frac{\hat{r}}{d_{HFPP}})^2 + 1}
\end{equation}

with the phase shift amplitude $\varphi_{HFPP}$~$<$~0 and the size $d_{HFPP}$. $d_{HFPP}$ and $\varphi_{HFPP}$ are difficult to control experimentally as they depend on several experimental parameters such as the ZOB size, the film temperature or the irradiation time \citep{VPP_2014,2018_Simon_NegCharge,2019_Simon_LN2_Charging}. Image simulations were conducted for various values for both parameters but $d_{HFPP}$~=~100~nm and $\varphi_{HFPP}$~=~-0.5$\uppi$ were used for the shown images and contrast analysis. The resulting profile is displayed in Figure~\ref{F:PPLineScans}a. The continuous thin film with a uniform assumed thickness of $t_{HFPP}$~=~8~nm adds a constant phase offset to the distribution, which is neglected in the phase profile but damps the amplitude of the transmitted wave following Eq.~\ref{Eq:AmpMFP}. Positive PC is observed in HFPP images.

\paragraph{Electrostatic Boersch PP}

The electrostatic Boersch (B)PP consists of a micro-structured device that serves to create a homogeneous phase shift within a central hole with radius $r_{BPP,in}$ of a ring electrode carried by supporting arms \citep{BoerschOptimize.2007}. The BPP induces a variable phase shift $\varphi_{BPP}$ for $\hat{r}<r_{BPP,in}$ and electrons are obstructed by the ring electrode ($r_{BPP,in} \leq \hat{r} < r_{BPP,out}$) and the supporting arms with a width of $d_{BPP,arm}$. In these regions, $A_{PP}$ is equal to 0 (black area in Figure~\ref{F:PPs}d). The values used for the presented simulations are $r_{BPP,in}$~=~750~nm,  $r_{BPP,out}$~=~2~$\upmu$m and $d_{BPP,arm}$~=~1.5~$\upmu$m. PC may be positive (negative applied phase shift by the BPP) or negative (positive applied phase shift).

\paragraph{Electrostatic Zach PP}

The electrostatic Zach PP was developed from the Boersch PP to minimize the obstruction of electrons \citep{PPZach_MM2010}. The Zach PP consists of a single rod with an open-ended electrode generating an inhomogeneous potential displayed in Figure~\ref{F:PPs}e. The rod end is positioned at a distance $d_{Zach}$ to the ZOB and the potential is scaled to reach the desired, adjustable phase shift at the position of the ZOB. Both the vertical (parallel to the rod) and horizontal phase profiles are plotted in Figure~\ref{F:PPLineScans}. $d_{Zach}$ was chosen to 1~$\upmu$m for the results presented in the main article being a value typically used in experiments \citep{Obermair2020}. PC may be positive (negative applied phase shift by the Zach PP) or negative (positive applied phase shift).

\paragraph{Tunable Ampere PP}

The tunable Ampere (TA)PP consists of a nano-scale current-carrying wire, which is placed close to the ZOB \citep{Marco_TunablePP_2018}. The wire contains a part oriented parallel to the electron-beam direction, which creates a phase-shifting field. This field is calculated as detailed in the supplementary material of \citep{Marco_TunablePP_2018} with a wire width of 200~nm, a length of the vertical part of 4~$\upmu$m and a variable applied current $I_{TAPP}$. The wire is placed at a distance $d_{TAPP}$ (350~nm for the shown images plots) to the ZOB. The vertical (parallel to the wire) and horizontal phase profiles are plotted in Figure~\ref{F:PPLineScans}a. PC in TAPP images is positive.

\paragraph{Laser PP}

The laser (L)PP is based on the interaction between a highly intense standing laser wave and the transmitting electrons \citep{Glaeser_LaserPP_2019}. In this study, the phase profile is calculated as detailed in the supplementary information of \citep{Glaeser_LaserPP_2019} with the identical parameters, being $\lambda_{L}$~=~1064~nm and NA~=~0.026. The PP is assumed being perfectly aligned with respect to the ZOB. The phase shift is depicted in Figure~\ref{F:PPs}g as well as the vertical and horizontal profiles in Figure~\ref{F:PPLineScans}a. The positive phase shift of the LPP induced on the ZOB leads to a negative PC in the images.

\paragraph{Foucault aperture}

In addition to active PPs, simple blocking structures have been used to enhance phase contrast in phase-contrast images. The Foucault aperture, obstructing half of the plane \citep{Cullis.1975}, is included in the present study and is positioned at a distance $d_{Ap}$, which was set to 350~nm.

\subsubsection{Assumed microscopes}
\label{S:Mics}
PPs are typically positioned in the back-focal plane (BFP) of a transmission electron microscope and the focal length $f_{BFP}$ together with the electron wavelength $\lambda$ determines the transformation from the real-space PP ($\hat{x},\hat{y}$) distributions to the reciprocal space $(u_{x},u_{y})$ required for calculation of the image formation process (Eq.~\ref{Eq:ImageFormation}):

\begin{equation}
    (u_{x},u_{y})= \frac{(\hat{x},\hat{y})}{\lambda f_{BFP}}
    \label{Eq:RealRecipSpace}
\end{equation}

Two microscopes equipped with field-emission electron guns were assumed for the conducted simulations, whose parameters are detailed in Table~\ref{T:Mics}. For the study of the bacteriophages and the aC NPs, a microscope with an electron energy of 300~keV (named \textit{cryo-Mic}) similar to the one used in a recent experimental study is assumed \citep{Obermair2020}. In addition to the regular BFP ($f_{BFP}$~=~5~mm), the microscope may be readjusted to obtain a magnified (M)BFP with significantly larger $f_{MBFP}$~= 15.8~mm. The thin-film based PPs, the TAPP and the blocking apertures are placed in the BFP, while the electrostatic Boersch and Zach PPs are positioned in the MBFP as it has already been reported experimentally \citep{ZPP_Barton_Boersch_UM2011,Obermair2020}. 

For the graphene simulations, all PPs with exception of the LPP are placed in the regular BFP ($f_{BFP}$~=~1.8~mm) of a microscope with high-resolution pole-piece equipped with an aberration corrector (named \textit{HR-Mic}) and operated at 80~keV. For the LPP, the focal length is changed to $f_{BFP,laser}$~=~20~mm for the NPs and graphene simulations to reflect the value achieved in the experimental study \citep{Glaeser_LaserPP_2019} without modifying the other experimental parameters. For the phages, $f_{BFP,laser}$ is lowered to 17.5~mm to obtain an integer sampling (12 px per node) of the strongly varying phase shift of the laser field in horizontal direction. Many PP designs need a supporting structure and are fabricated in an aperture. To obtain a similar result for all PP images, the objective aperture was set to a fixed spatial frequency.

\begin{table}[]
    \centering
    \begin{tabular}{|c|c|c|c|}
    \hline
 & \multicolumn{2}{c|}{\textit{cryo-Mic}} & \textit{HR-Mic} \\
 & NPs & Phages & \\
 \hline
 $\lambda$ / pm & 1.97 & 1.97 & 4.18 \\
 $\Delta$ / nm  & 8 & 8 & 4.5\\
 $\alpha$ / mrad   & 0.1 & 0.1 & 0.3 \\
 $Z$ / nm & -70 & -70 & -5 to +5 \\
 $C_S$ / mm  & 5 & 5 & 0.001\\
 $u_{OA}$ / nm\textsuperscript{-1} & 2.3, 4.65, 9.3 & 3 & 12 \\
  $f_{BFP}$ / mm & 5 & 5 & 1.8 \\
 $f_{MBFP}$ / mm & 15.8 & 15.8 & -\\
 $f_{BFP,laser}$ / mm  & 20 & 17.5 & 20 \\
  \hline
    \end{tabular}
    \caption{Parameters for the assumed microscopes with the focal length of the regular BFP $f_{BFP}$, the magnified BFP $f_{MBFP}$ and BFP assumed for the LPP $f_{BFP,laser}$. Values are given for the defocus $Z$, the spherical aberration constant $C_S$, the focal spread $\Delta$, the semi-convergence angle $\alpha$ and the objective aperture $u_{OA}$. $u_{OA}$ of the \textit{cryo-Mic} is given for the three studied wave-functions with NP diameters 15-20 nm, 6-12.5 nm and 1-5 nm. }
    \label{T:Mics}
\end{table}

\subsubsection{Cut-on frequency}

An important characteristic of a PP design is its cut-on frequency, which indicates from which spatial frequency on the PP is effective. It is a clear definition for hard-edge PPs such as the ZPP, where it corresponds to the radius of the central hole as electrons passing through the hole are not phase shifted. The same holds true for the HPP, BPP or Foucault aperture where the respective radius or distance to the ZOB defines the cut-on frequency. The cut-on frequency for the PPs with a hard edge positioned at a distance of 350~nm to the ZOB in the BFP of the \textit{cryo-Mic} is 0.036~nm\textsuperscript{-1}, which corresponds to an object size of 28~nm. The outer radius of the BPP positioned in the MBFP leads to a cut-on frequency of 0.06~nm\textsuperscript{-1} (object size of 16~nm). In the \textit{HR-Mic}, these cut-on frequencies increase to 0.097~nm\textsuperscript{-1} (ZPP, HPP, Foucault) and 0.53~nm\textsuperscript{-1} (BPP) due to the reduced focal length.

For PPs with smoothly varying phase profile, the cut-on frequency is not given by a single frequency but rather by a range of spatial frequencies over which the induced phase shift changes. Figure~\ref{F:PPLineScans}a shows the comparison of the phase profiles of the PPs with a smooth profile for the \textit{cryo-Mic}.

\subsubsection{Aspects of pixel-based simulations}

When generating a suitable OEWF for image simulations, the objects need to be rendered with a sufficient amount of pixels to avoid the introduction of artifacts. Placing objects too close to the border of the OEWF may cause unwanted interference with the latter. When PPs are considered, sampling of the reciprocal space (i.e. of the BFP) needs additionally to be considered. While an OEWF, in which the object is rendered with a high amount of pixels may suffice for CTEM simulations, the sampling of the reciprocal space may still not be sufficient to obtain a correct representation of the PP phase profiles. The pixel size in the BFP $\hat{s}_{px}$ depends on the used pixel size of the OEWF in real space $s_{px}$ and the total number of pixels in the OEWF $n_{px}$:

\begin{equation}
    \hat{s}_{px} = \frac{1}{s_{px} n_{px}}
    \label{Eq:BFPSampling}
\end{equation}

For small pixel sizes $s_{px}$, the OEWF size may need to be increased drastically to yield a good representation of the PP profiles. In the present study, the OEWF size was 4096~px for the graphene ($s_{px}$~=~13~pm) and NP samples ($s_{px}$~=~54, 108 and 217~pm), while 2048~px were chosen for the phages ($s_{px}$~=~190~pm).

\subsubsection{Finite zero-order beam diameter}

In CTEM imaging without PP, the effect of partial spatial coherence, stemming from the finite source size, can be considered by an envelope function (Eq.~\ref{Eq:EnvelSpatialCoherence}) \citep{ReimerTEM_5}. The finite source size leads to electrons that do not fulfill the parallel-beam condition but travel at a small angle with respect to the optical axis. The finite source size leads to a finite size of the ZOB, which may be assumed as 2D Gaussian distribution. In contrast to CTEM, the phase and amplitude induced by PPs may vary strongly around the ZOB leading to a considerable change of the image contrast already for small variations of the exact ZOB position. To consider this effect, the final image is calculated as weighted sum of individual images with shifted PP profiles. The number of images contributing to the sum is determined by the size of the ZOB $\sigma_{ZOB}$ and the pixel size in the BFP. All pixels within a circle around the central pixel (representing the ZOB) with a diameter of 2$\sigma_{ZOB}$ + 1px are considered as visualized in Figure~\ref{F:PPLineScans}b. The weighting factors are then calculated by normalizing the sum of all contributing pixels using a 2D Gaussian function.

For the sake of comparability with PP dimensions, we give $\sigma_{ZOB}$ by its real-space size in the BFP ($\hat{x},\hat{y}$), which needs to be scaled for the enlarged BFPs. In addition, we indicate the corresponding convergence angle caused by the finite source size and the illumination settings for orientation. A finite ZOB of $\sigma_{ZOB}$~=~50~nm (0.1 mrad) has been assumed for the simulation of the images of the T4 sample, while it was not considered for the aC NPs and graphene samples. In case of the aC NPs, the effect of an increasing ZOB was studied separately (section~\ref{S:FiniteZOB}). Due to the small feature size in the graphene sample, the effect may well be neglected for the corresponding simulations.

\subsection{Image evaluation criteria}

To obtain a quantitative and comparative evaluation of the simulated images, contrast $C$ and visibility $V$ are determined from the mean ($\overline{I}$) and minimum and maximum ($I_{min}$,$I_{max}$) intensities in the object (O) and background (BG) areas:

\begin{align}
    C &= \frac{\overline{I}_O - \overline{I}_{BG}}{\overline{I}_{BG}} \\
    V &= \frac{I_{O,max} - I_{O,min}}{\overline{I}_{BG}}
\end{align}

Background areas are chosen from object-free regions with sizes sufficiently large to average out any noise and away from possible shadow images or halos introduced by the PPs. The object areas for the NP samples correspond to circles encompassing the entire NP and in case of the phages, four rectangular object regions are defined selecting the cylindrical bodies and an element of the tail of each of the phages as marked in Figure~\ref{F:OEWFs}a. In case of the graphene sample, a single C atom with a diameter of 80~pm is selected as object. In this definition, a positive value of contrast C corresponds to negative PC and vice versa.

\newpage

\section{Comparison of different PP approaches}

Based on the image formation process with PPs, a custom Matlab based software with graphical user interface was developed from the one used in recent publications \citep{Obermair2020,Obermair2021GradedZPP}. The software is described in the SI (section \ref{SIS9}) and has been published and is free to use\citep{PPSim2}. This Section contains the results from the conducted image simulations and their discussion. 

\subsection{T4 Bacteriophages}

\begin{figure}[ht]
    \centering
    \includegraphics[width=\linewidth]{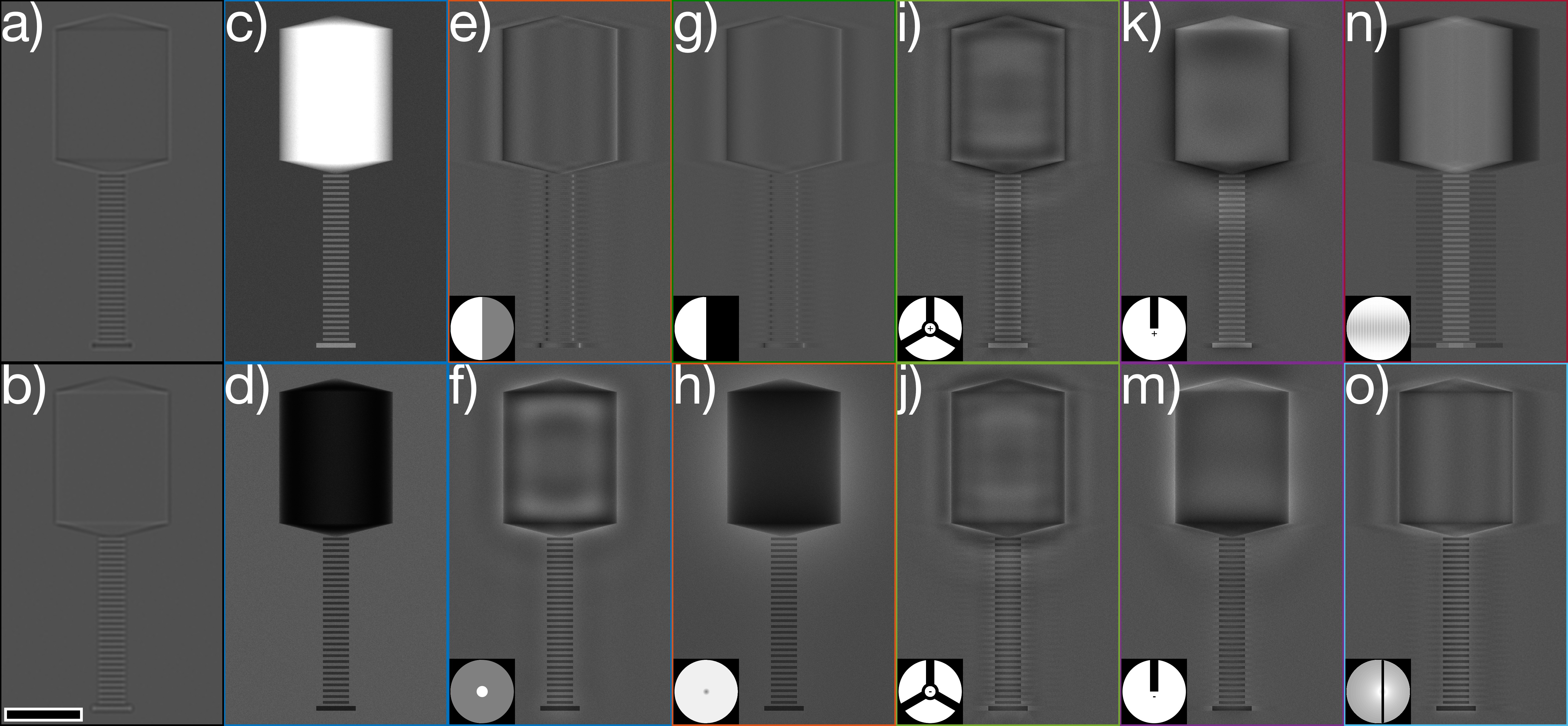}
    \caption{Simulated images of the vertically oriented phage for (a,b) CTEM at $Z$~=~$\mp$3.5~$\upmu$m, (c,d) ideal PP, (e,f) HPP and ZPP, (g) Foucault aperture, (h) HFPP, (i,j) BPP and (k,m) Zach PP, (n) LPP and (o) the TAPP with $I_{TAPP}$~=~3.5~mA. Phase shift of PPs with variable phase shift was set to $\pm\uppi/2$, respectively. The calculations were performed with a ZOB size of 50~nm (0.1 mrad) and the intensity ranges from 0.1 (black) to 3 (white). Image intensity was clipped at a value of 3 to improve visibility of images acquired with non-ideal PPs. Scale bar is 50~nm.}
    \label{F:T4Images}
\end{figure}

Figure~\ref{F:T4Images} shows exemplary images of a vertically oriented T4 bacteriophage simulated considering different PPs and conventional (C)TEM. The phage, barely visible in the CTEM image in under- or overfocus (Fig.~\ref{F:T4Images}a, b) appears with strong contrast in images with ideally assumed PP (Fig.~\ref{F:T4Images}c,d). In the images obtained with a HPP or Foucault aperture, no actual PC is observed but phase jumps in the object located at the borders of body and tails and oriented parallel to the mask edge are revealed (Fig.~\ref{F:T4Images}e,g). Opposing edges thereby appear in inverted contrast. While the general appearance is similar, the Hilbert PP leads to an increased magnitude of the topographical, or differential contrast (Fig.~\ref{F:T4Images}e). In case of the ZPP, the tail appears with positive PC while fringing artifacts are dominant for the body, which only shows slight positive PC (Fig.~\ref{F:T4Images}f). Significant positive PC is observed for the HFPP (Fig.~\ref{F:T4Images}h), for which the phage body is surrounded by a bright halo, the typical artifact observed in HFPP images \citep{Pretzsch.2019,Obermair2020}.

\begin{figure}[ht]
    \centering
    \includegraphics[width=0.67\linewidth]{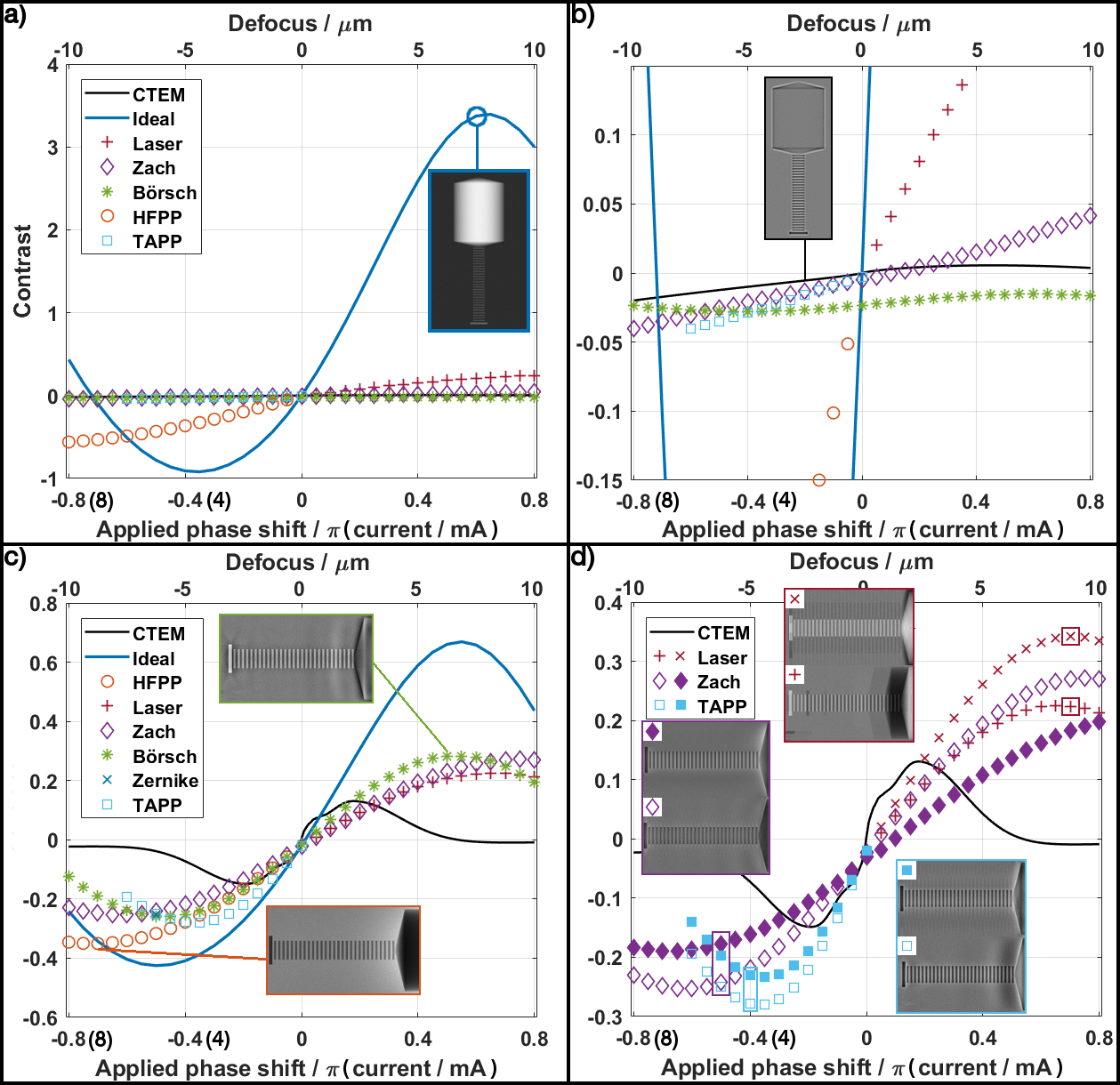}
    \caption{Contrast analysis of the T4 phage. (a) At full scale, only the ideal, HFPP and LPP cause a PC of the horizontally oriented phage body different from zero. (b) The magnified scale shows that the Zach and TAPP create minor PC, while the BPP mainly causes obstruction-induced contrast. (c) Comparable contrast values of the horizontally oriented tail are obtained for the different PPs exceeding the maximum contrast achievable by CTEM and getting closer to the ideal PP case. (d) Asymmetric phase profiles cause different PC of the horizontally (+, hollow diamond and hollow square markers) and vertically (x, filled diamond and filled square markers) oriented tail.}
    \label{F:T4Contrast}
\end{figure}

While the repetitions of the tail of the phage may be well discerned for the different electrostatic PPs, the phage body shows negligible PC but pronounced fringing artifacts for the BPP (Fig.~\ref{F:T4Images}i,j). The fact that inverting the phase shift only marginally changes the PC of the body indicates that the contrast mainly arises from obstruction by the BPP supporting structure, as visible also in a simulated image only considering this supporting structure (see SI, Figure \ref{SF:T4_Boersch}). An inversion of the PC with the applied phase shift is observed for the Zach PP although the contrast is weaker than in the ideal or HFPP case (Fig.~\ref{F:T4Images}k,m). The asymmetric Zach profile manifests itself at the borders of the phage body. Fringing artifacts as well are visible within the phage body in case of the TAPP (Fig.~\ref{F:T4Images}o). For the LPP, a strong negative PC may be observed, which, however, is overlayed with shadow images stemming from the oscillations of the laser field (Fig.~\ref{F:T4Images}n).  Additional images of the T4 are shown in the SI (section \ref{SIS4}) and illustrate the dependence of the achievable contrast on phase shift and cut-on frequency of the different PPs.

Figure~\ref{F:T4Contrast} shows the contrast analysis of the horizontally oriented phage body (Fig.~\ref{F:T4Contrast}a, b) and tail (Fig.~\ref{F:T4Contrast}c and d). Figure~\ref{F:T4Contrast}a reveals the strong contrast achievable with an ideal PP. The maximum PC is achieved for a phase shift different to $\pm$0.5~$\uppi$ as the body alters the phase of the electron wave considerably and may not be considered a WPO. At this scale, only the HFPP and LPP yield a PC noticeably differing from zero.

The plot at magnified scale (Figure~\ref{F:T4Contrast}b) reveals a weak, controllable contrast in the Zach PP and TAPP images, while the BPP leads to a negative contrast only slightly varying with the phase shift. The Zernike PP (not plotted) shows a value comparable to the TAPP at an applied current of 5~mA. It is noted that all of the PP designs with variable phase shift, except the BPP, reach maximum PC at the highest studied phase shift signifying that a higher contrast would be possible at higher applied phase shifts. The reason is that the effective phase shift between the ZOB and the spatial frequency related to the diameter of the phage body (see green vertical bar in Figure~\ref{F:PPLineScans}a) is lower than the actually applied phase shift at the position of the ZOB. This is due to the spatial extension of the phase-shifting profiles towards higher spatial frequencies leading to a non-zero induced phase shift at the spatial frequency of the phage body.

The contrast analysis of the tail of the horizontally oriented phage is depicted in Figure~\ref{F:T4Contrast}c. In this case, the contrast gain achievable by PPs in comparison with CTEM is lower than for the body, but still significant. The different PP designs now approach the ideal PP case with the HFPP getting the closest. The displacement of the maximum contrast to an applied phase shift of -0.7$\uppi$ for the HFPP is caused by the considered finite ZOB size ($\sigma_{ZOB}$~=~50~nm), which effectively lowers the induced phase shift (see Section~\ref{S:FiniteZOB} and \ref{SIS6}). The electrostatic PPs allow adjusting the contrast between positive and negative PC and yield similar contrast values, which are also comparable to the ZPP.

The orientation dependence of the tail contrast is analyzed for the LPP, Zach PP and TAPP in Figure~\ref{F:T4Contrast}d by comparison of the tail contrast in dependence of the phage orientation. The Zach and LPP show considerable differences in tail contrast due to their inhomogeneous phase profiles, while the difference for the TAPP is not as pronounced. Due to the presence of the supporting rods, the BPP as well causes an orientation-dependent contrast (see SI, Section \ref{SIS2}).

\subsection{aC nanoparticles embedded in ice}

\begin{figure}
    \centering
    \includegraphics[width=\linewidth]{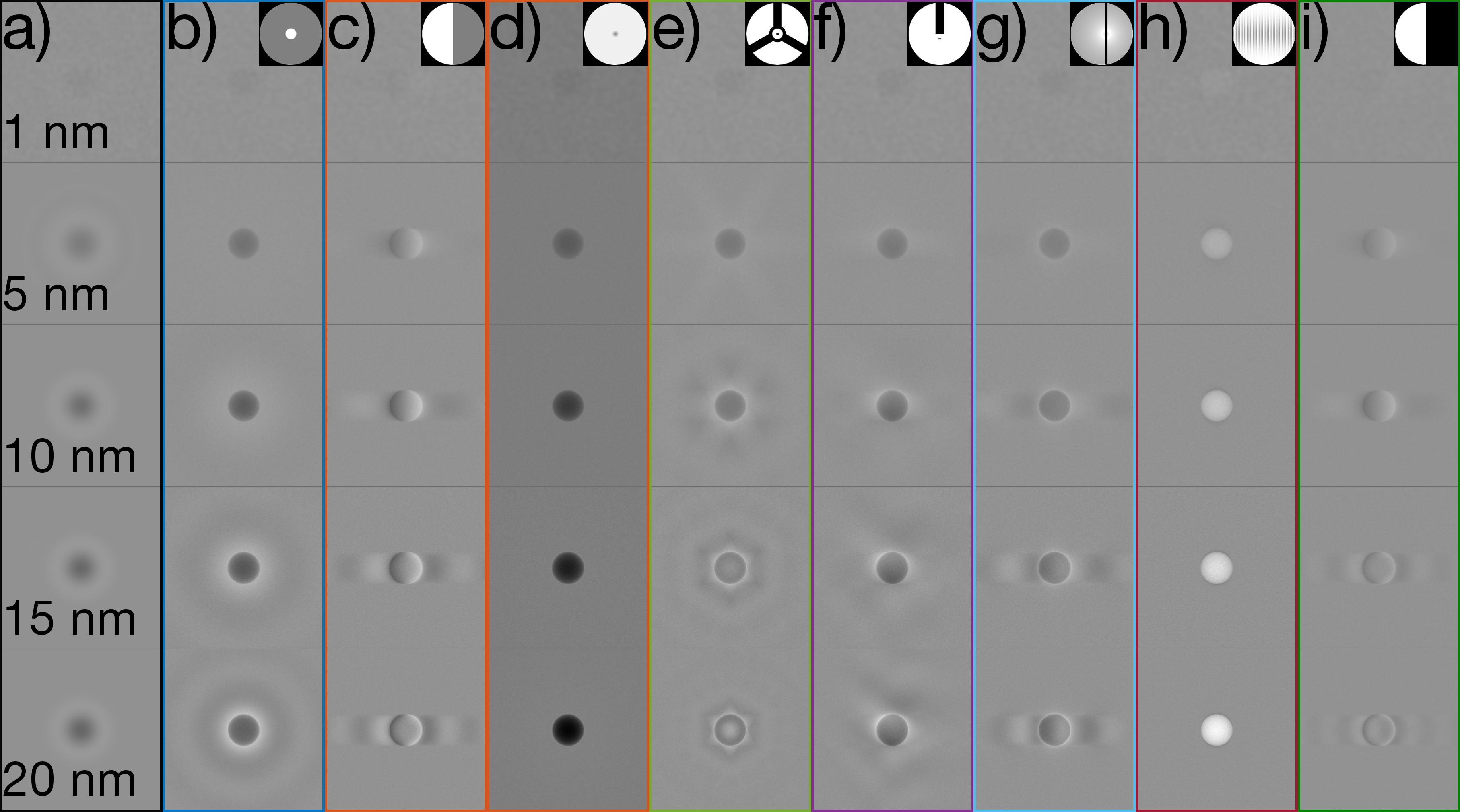}
    \caption{Comparison of simulated images of spherical aC NPs embedded in vitrified ice with each row of images corresponding to the same diameter between 1 and 20~nm. The images are calculated for (a) defocused CTEM (see text and SI for values), (b) ZPP, (c) HPP, (d) HFPP, (e) BPP, (f) Zach PP, (g) TAPP at 3~mA, (h) LPP and (i) obstructing Foucault aperture. The applied phase shift was set to $\pm$+-pi/2 for PPs with variable phase shift. The images were calculated without considering a finite ZOB.}
    \label{F:Spheres}
\end{figure}

\begin{figure}
    \centering
    \includegraphics[width=\linewidth]{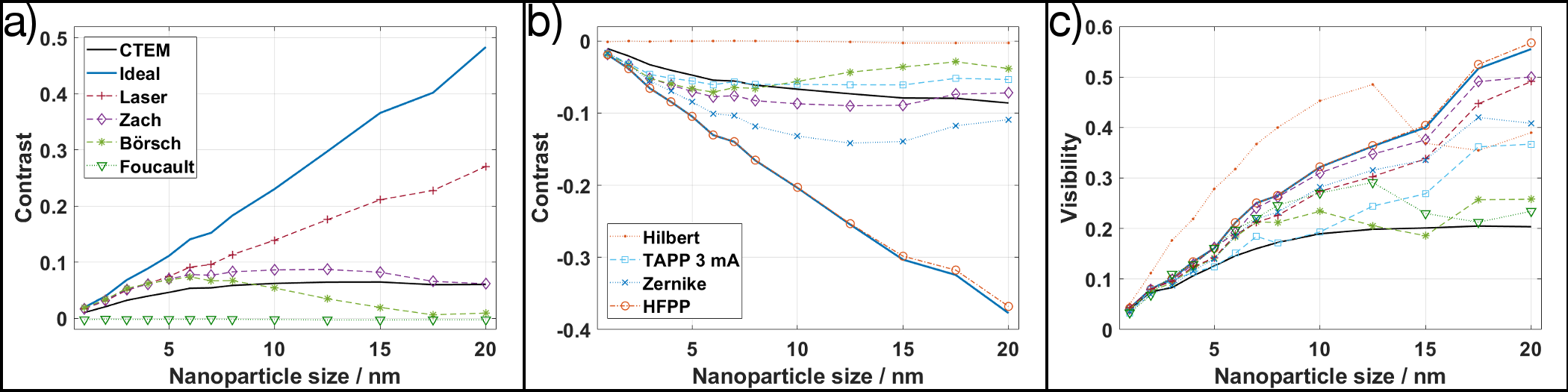}
    \caption{Contrast of spherical aC NPs embedded in 30~nm of vitrified ice achieved by PPs generating (a) negative PC and (b) positive PC depending on the NP diameter and (c) visibility for all studied PPs.}
    \label{F:SpheresCon}
\end{figure}

Images of spherical aC NPs with 5 different diameters embedded in vitrified ice with a thickness of 30~nm are shown in Figure~\ref{F:Spheres}. To determine the contrast reachable in CTEM, images for a range of $Z$ values of up to $\pm$150~$\upmu$m for the largest NPs were simulated and the $Z$ values generating maximum CTEM contrast were used for Figures~\ref{F:Spheres} and \ref{F:SpheresCon}. The $Z$ values lie between 450~nm and 57~$\upmu$m for the NP with a diameter of 1 and 20~nm, respectively. The results are discussed in the SI (section \ref{SIS3}). The PP simulations were conducted without considering a finite ZOB, whose effect is analyzed in Section \ref{S:FiniteZOB}.

The NP with a size of 1~nm may be discerned with minor contrast for CTEM and PP images (Fig.~\ref{F:Spheres}a-h) but is hardly visible for the obstructing aperture (Fig.~\ref{F:Spheres}i). The 5~nm NP appears clearly visible against the background in all PP images and only negligible artifacts, mainly manifesting themselves as characteristic halos around the NPs, may be observed. For example, the BPP leads to a star-like halo around the NP, while a bright (dark-bright), horizontally oriented halo is visible for Zach PP (TAPP), respectively. The HPP and obstructing apertures lead to differential contrast perpendicular to the PP edges, which is stronger for the HPP in comparison with the obstructing aperture. With increasing NP size, these PP artifacts gain importance, fringing for hard-edge PPs strongly increases and halos become more pronounced for smooth and inhomogeneous phase profiles. The PP that is least affected by artifacts is the HFPP, where only a minor halo around the 20~nm NP may be observed and the overall intensity in the image is reduced due to amplitude damping in the film. Strong PC is generated by the LPP, which causes only a slightly dark halo around the NPs. Shadow images are also observed for the LPP but they are located outside of the shown area (see Fig. \ref{SF:Sp_Laser}).

Contrast and visibility of the NPs depending on their diameter is shown in Figure~\ref{F:SpheresCon}. PPs generating negative PC and the obstructing aperture are displayed in Figure~\ref{F:SpheresCon}a. As the differential contrast caused by the Foucault aperture averages out when determining the contrast of the entire NP, the NPs show zero contrast for all diameters. The contrast achieved by the PPs follows the ideal contrast for low NP sizes but starts to deviate considerably already for NPs with a diameter of 3~nm. The LPP shows the highest contrast throughout the range followed by the Zach PP, while the BPP falls below the contrast obtainable from defocused CTEM for sizes above 10~nm.

Figure~\ref{F:SpheresCon}b reveals the contrast of the PP designs causing positive PC and the HPP. Similar to the obstructing aperture, the contrast of the HPP is always zero. The dependence of Zach and BPP on the NP diameter is comparable to the negative PC case (compare Fig.~\ref{F:SpheresCon}a with \ref{F:SpheresCon}b) and the TAPP yields a contrast that lies in between these two designs. The contrast achieved by the ZPP is considerably larger while the HFPP is identical to the ideal case for sizes below 15~nm and only slightly differs for NPs up to 20~nm diameter.

The visibility displayed in Figure~\ref{F:SpheresCon}c shows the potential of the HPP, whose differential contrast leads to a visibility superior to the ideal case up to sizes of 15~nm. The obstructing aperture only yields approximately half of the HPP visibility values. The PC, in addition to fringing and halo artifacts introduced by the rest of the PPs, leads to a visibility that is close to the ideal case over the whole investigated range of sizes.

It is worth noting that contrast and visibility plotted for CTEM represent the maximum obtainable values, which are reached at a specific $Z$ for each NP diameter (SI, Section \ref{SIS3}). This signifies that in a CTEM image, where, e.g., the NP with a diameter of 20~nm reaches its maximum contrast ($Z$~=~57~$\upmu$m), the contrast of any other NP will be significantly lower than the values indicated in Figure~\ref{F:SpheresCon}. This is in contrast to the PP images, where the data corresponds to a single set of values and varying parameters may improve the reachable contrast and visibility. Among these parameters are the defocus $Z$, the cut-on frequency of hard-edge PPs and the induced phase shift for variable phase shift PPs. Especially the PPs that combine a variable phase shift with a smooth phase profile, which can be deliberately positioned in the BFP, provide enhanced control and flexibility regarding the achievable contrast (section \ref{SIS4}).

\subsection{Graphene}

\begin{figure}
    \centering
    \includegraphics[width=0.8\linewidth]{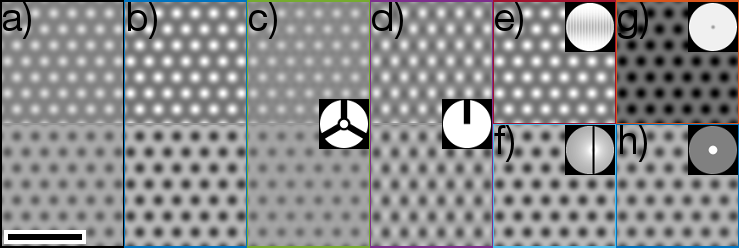}
    \caption{Images of a graphene monolayer for (a) CTEM with $Z$~=~$\pm$5~nm. PP images at $Z$~=~0~nm for the (b) ideal PP ($\pm\uppi$/2), (c) BPP ($\pm\uppi$/2), (d) Zach PP at a distance of 2~$\upmu$m ($\pm\uppi$/2), (e) LPP ($\uppi$/2), (f) TAPP at 2~mA, (g) HFPP (-$\uppi$/2) and (h) ZPP.  The ZOB was assumed as single pixel. The scale bar is 1~nm.}
    \label{F:Graphene}
\end{figure}

Images obtained from image simulations with the graphene monolayer sample are shown in Figure~\ref{F:Graphene}. In all cases, the graphene lattice is resolved at $Z$~=~0~nm, with only minor artifacts or orientation-dependent contrast visible in the atomic columns. The contrast achieved by implementation of any PP (Fig.~\ref{F:Graphene}b-h) is comparable with CTEM for a defocus of $\pm$5~nm (Fig.~\ref{F:Graphene}a). 

\begin{figure}
    \centering
    \includegraphics[width=0.66\linewidth]{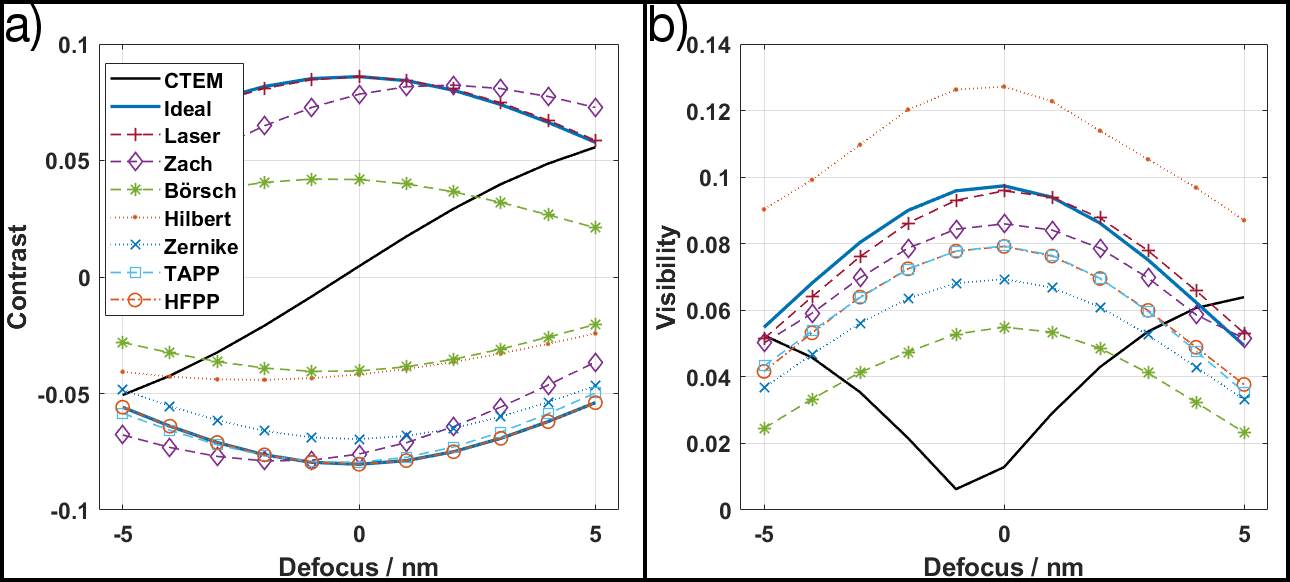}
    \caption{(a) Contrast and (b) visibility of a single C atom in the graphene matrix in dependence of $Z$ for CTEM and various PPs.}
    \label{F:GraphenePlot}
\end{figure}

The advantage of PPs in TEM at atomic resolution is revealed by Figure~\ref{F:GraphenePlot}. The contrast of a single C atom in the graphene matrix maintains its sign throughout the whole investigated defocus range between -5 and 5~nm if a PP is used. The reachable contrast is comparable for the displayed PP designs except for the BPP, which yields only a reduced contrast due to the large amount of obstructing matter present in the BFP. The plot of the visibility shows that the highest value is reached for $Z$~=~0~nm in case of a PP, where it reaches a minimum for CTEM. Additional images of the graphene sample for the whole studied defocus range may be found in the SI (section \ref{SIS5}).

\subsection{Discussion}

The comparative simulations show that an ideal PP improves phase contrast and image interpretability for a phase object of any size. The performance of the different PP designs is closely linked to their spatial phase distribution and possible obstructing elements. The phase profile has to be narrow to allow contrast enhancement for larger objects. Obstructing elements and sharp edges in the phase profile cause fringing in the images, while halos are generated by smooth gradients in the phase profile. Additionally, the suitability of a specific PP depends on the object to be investigated. While objects with sizes smaller than 5~nm may be imaged with high phase contrast by basically any of the studied PPs, the performance starts to decrease considerably for objects with larger sizes depending on the PP type and its cut-on frequency and spatial phase-shift distribution. In contrast to obstructing material, the damping effect of transparent material does only lead to an overall reduction of the image intensity as both inelastically and elastically scattered electrons in the PP are taken out of the image formation process, but does not hamper the reachable contrast in the simulated images. In experiments, parts of the elastic scattering will contribute to the background intensity and thus slightly lower the contrast. Moreover, these scattered electrons contribute to the dose the sample is exposed to and lower the dose-limited resolution of beam-sensitive objects.

From the studied experimental realizations, the HFPP shows a performance getting the closest to the ideal case with the only artifact of a bright halo visible around large objects, which starts for the used settings from a size of 20~nm. These beneficial phase-shifting properties, together with a relatively easy implementation, has led to numerous applications mainly in cryo-TEM but also in the field of material science \citep{PPReviewMalacHettler}. However, the presented results show that a consideration of the HFPP as ideal PP may be erroneous in terms of a quantitative interpretation. Quantitative imaging as well requires an exact knowledge of shape and amount of the induced phase shift, which are difficult to control experimentally in case of the HFPP. Regarding the imaging of crystalline samples, the formation of additional phase-shifting patches on the HFPP film induced by the intense diffracted beams has to be considered \citep{HaradaGrating_2020}. This effect leads to a changing contrast with the irradiation dose, making the HFPP not ideally suited for applications in material science concerned with crystalline objects.

Among the other studied PPs, the LPP generates strong PC for all object sizes, but shadow images are generated by the oscillating laser field, which interfere with the same object (Fig.~\ref{F:T4Images}n) or appear at a certain distance of it (Fig.~S14). Strong PC is also observed for the Zach, ZPP and TAPP although halos or fringing artifacts start to be strongly visible for object sizes larger than 10~nm. The performance of a Zernike PP depending on the cut-on frequency and phase shift has been subject to in-depth studies earlier \citep{ZPPOptimizing_NagayamaDanev_UM2011}. A beneficial application of the ZPP, Zach PP and HFPP on the imaging of T4 bacteriophages has been demonstrated experimentally \citep{ZPP_cryo_tomo_Rado_2010,Obermair2020}, showing comparable results between HFPP and the Zach PP implemented in a MBFP. The BPP performs worst of all the studied active PPs as artifacts dominate the image appearance starting from an object size of 10~nm and yielding the lowest contrast. The HPP is useful to make phase objects visible by means of a differential contrast, which is also created by a Foucault aperture. The latter however causes less visibility than the HPP. 

Hardware design optimizations with the aim to reduce PP artifacts and improve their phase-shifting behavior have been presented for various PPs, e.g. electrostatic devices that lie between the BPP and Zach PP \citep{BoerschPP_UM2015_Walter} or an additionally structured ZPP \citep{ZPPOptimDesign2016}. These optimizations will shift the achievable contrast.

The simulations of the bacteriophages show the advantage of a variable phase shift as the typically chosen phase shift of $\pi$/2 suitable to maximize contrast of WPOs is not the optimum for larger, strong phase objects. For PPs with a smooth phase-shift distribution, a variable phase shift allows to additionally tune the cut-on frequency. A variable phase shift is as well essential for PP based object-wave reconstruction \citep{Gamm.2010}.

The simulations of the graphene lattice at atomic resolution show that high-resolution (HR)TEM could profit from the application of PPs as the contrast of phase objects can be maintained positive or negative over a larger focus range leading to an improved image interpretability. This is particularly interesting for 2D materials, which are known to show buckling, which leads to contrast variations in HRTEM \citep{Meyer.2007,Butz.2014}. Previous studies showed that, in addition, the total reachable contrast is increased if PPs are employed \citep{Gamm.2008EffectPP}, which could facilitate an easier detection of dopants with similar atomic number than the host (e.g., N in graphene or in single-walled carbon nanotubes \citep{Ayala.2010,Arenal.2014}) in comparison to CTEM. Of particular interest in HRTEM is the application of object-wave reconstruction, which would allow to obtain the ultimate information achievable from a sample, its OEWF.

Summarizing, the simulations show that a PP, that can be kept clean from contamination and charging effects, can provide significant improvement of phase contrast or image interpretability for a wide range of different samples. As PPs allow the in-focus imaging at high contrast, delocalization and fringing effects induced by defocused CTEM can be avoided.

\section{Effect of different simulation parameters}
\label{S:PPPar}

Numerous parameters have to be considered to obtain a realistic TEM image simulation. Three important parameters that are of special importance to PP images are discussed in this Section.

\subsection{Finite size of zero-order beam}
\label{S:FiniteZOB}

\begin{figure}
    \centering
    \includegraphics[width=\linewidth]{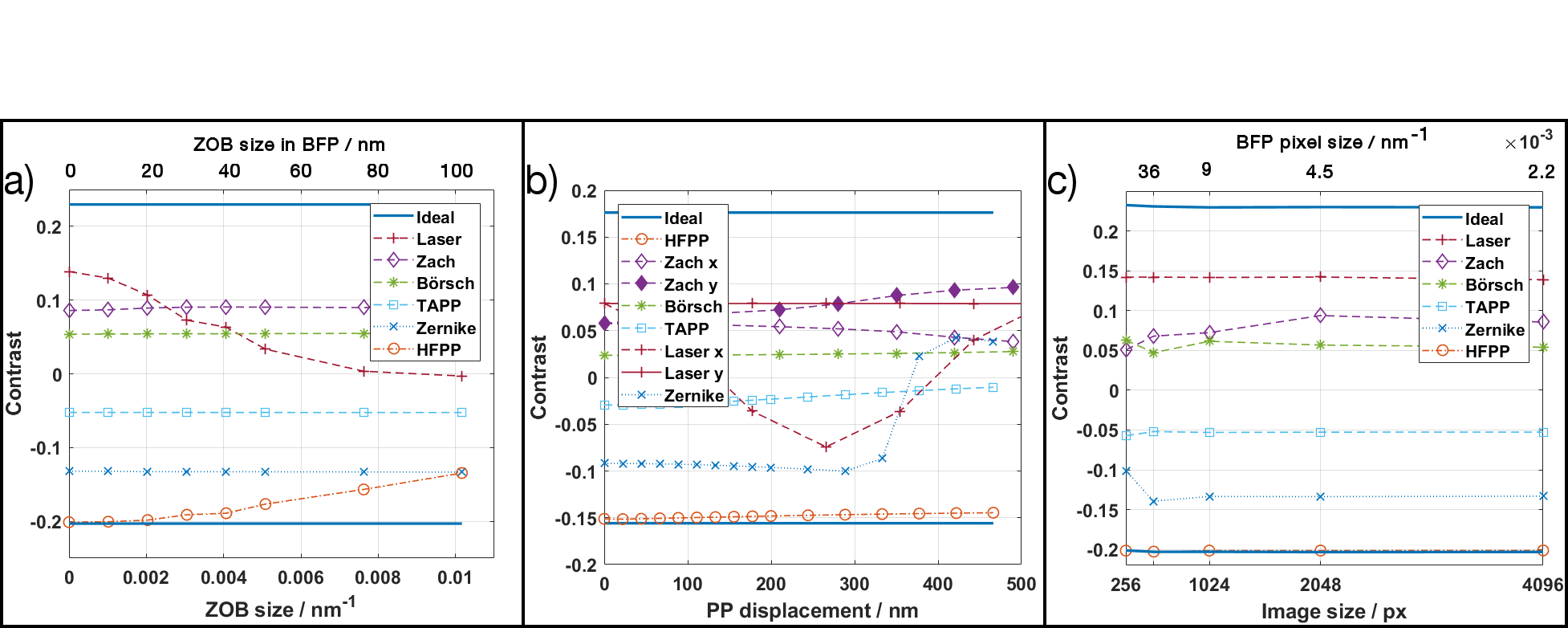}
    \caption{(a) Contrast of a 10~nm aC NP embedded in vitrified ice in dependence of the ZOB size $\sigma_{ZOB}$ for different PPs. $\sigma_{ZOB}$~=~100 nm corresponds to a convergence angle of 0.2 mrad. The second x axis showing the ZOB size in nm is given for the conventional BFP and is only true for the TAPP, the ZPP and HFPP and has to be scaled by the focal length for the MBFP (Zach, BPP) and laser BFP. (b) Contrast evolution of a 10~nm aC NP embedded in vitrified ice upon deliberate displacement of the PPs. In case of the HFPP, the displacement corresponds to a drift and yields an elongated phase profile. For the asymmetric phase profiles of Zach and LPP, the displacement is performed in two directions. (c) Contrast of a 5~nm aC NP embedded in vitrified ice depending on the total image size in pixel $n_{px}$ or the pixel size in the BFP. }
    \label{F:Params}
\end{figure}

The contrast of a 10~nm sized aC NP embedded in vitrified ice is shown for different PPs in dependence of the ZOB size $\sigma_{ZOB}$ in Figure~\ref{F:Params}a. $\sigma_{ZOB}$ is indicated in spatial frequencies as well as in real-space distance in the regular BFP ($\sigma_{ZOB}$~=~100 nm is equal to a convergence angle of 0.2 mrad). The corresponding size in the MBFP and laser BFP is larger and is obtained by scaling with the focal length. While the contrast for most PPs only changes marginally with an increasing $\sigma_{ZOB}$, the contrast goes down strongly for the HFPP and especially the LPP. This is due to their narrow phase profiles, which vary considerably within the ZOB size and thus leads to a reduction of the average induced phase shift if $\sigma_{ZOB}$ increases.

Although the contrast remains similar for the other PPs, an effect in image appearance can clearly be recognized. In general, a finite size of the ZOB leads to a smoothing of the image appearance, which also causes a reduction of fringing artifacts in sharp-edge PPs (section \ref{SIS6}). 

It is not straightforward to determine the experimental ZOB size in an experiment. Reported ZOB sizes are typically small (25~nm in \citep{VPP_2014}, 20-100~nm \citep{PPMikePractical_2016}) but large sizes have been reported (200-500~nm \citep{HFPP_2012}), especially in MBFPs (300~nm \citep{Obermair2020}). The size is mainly determined by the semi-convergence angle and may be controlled using the condenser aperture \citep{Pretzsch.2019}. The size of the HFPP distribution is linked to the ZOB size as it describes the beam that impinges on the HFPP film and creates the phase-shifting patch. However, these diameters, typically measured in diffraction mode, are most probably larger than the ZOB size that has to be considered for the image simulation, as the camera only "sees" a part of the illuminated sample area. The angular range of electrons contributing to the acquired image thus will be reduced in comparison to the measured diameter. Nevertheless, the presented results show that it is a non-negligible effect. The consideration of a finite ZOB size was important to obtain a quantitative or semi-quantitative link between simulated and experimental HFPP and Zach PP image appearance in recent studies \citep{Pretzsch.2019,Obermair2020,Obermair2021GradedZPP} and it explains the discrepancy between fringe intensity in experimental and simulated ZPP images not considering a finite ZOB \citep{Fukuda.2009}.

\subsection{PP misalignment}
\label{S:PPMis}

Except for the HFPP, where the phase profile is generated by the ZOB itself, all PPs need to be aligned with the electron beam. Figure~\ref{F:Params}b shows the contrast evolution of a 10~nm aC NP embedded in vitrified ice, when the different PPs are deliberately displaced from a centered position. In case of the HFPP, this displacement is considered by an elongated phase profile, as it would be caused by a drift of the PP prior to acquisition during the buildup of the phase-shifting patch. The results show that the NP contrast for the HFPP is only slightly reduced. However, an inhomogeneous halo appears around the NPs (section \ref{SIS7}), which has been shown to coincide with the experiment \citep{Pretzsch.2019}. 

The contrast remains similar for a slightly displaced ZPP and only the fringing becomes asymmetrically around the object (section~\ref{SIS6}), which has been observed experimentally \citep{Obermair2021GradedZPP}. However, once the ZOB leaves the central hole, the contrast is reduced drastically. No change in the contrast and a slight asymmetry in the halo is observed for the BPP as the studied displacement range is smaller than the inner ring radius. Minor changes are visible for the TAPP displaced along the wire. A difference depending on the direction of the displacement is observed for the Zach PP and the LPP due to their asymmetric phase profiles. The strongest change upon displacement is seen for the LPP in horizontal direction caused by its strongly oscillating field.

\subsection{Sampling}
\label{S:Sampling}

Figure~\ref{F:Params}c shows the dependence of the contrast of a 5~nm aC NP embedded in vitrified ice on the total number of pixels considered in the OEWF ($n_{px}$ in Eq.~\ref{Eq:BFPSampling}). While the object sampling is identical in all images, the contrast varies considerably for a number of PPs. Only the PPs with a very narrow distribution show a similar contrast for all sizes (HFPP and LPP) but a comparison of the respective images still reveals changes (SI, Section \ref{SIS8}). Both PPs with a smoothly varying phase profile (Zach and TAPP) and PPs with a sharp edge (ZPP and BPP) require an image size of $n_{px}$~=~2048 ($\hat{s}_{px}$=0.0045~nm\textsuperscript{-1}) for a correct outcome of the image contrast and appearance.

\section{Summary}

The comparative simulations reveal the contrast increase that can be achieved by different phase plates (PPs) in transmission electron microscopy (TEM). PPs are commonly said to enhance the phase contrast at "low spatial frequencies", but the presented study shows that the range of spatial frequencies, and corresponding object sizes, that can be beneficially imaged strongly depends on the used PP implementation. While artifact-free imaging may be achieved by many PPs for sizes of 5~nm and below, artifacts introduced by several PPs start to dominate for larger sizes. The simulated images give a visual impression of artifacts that can be expected in experimental images using such PPs. The contrast analysis reveals that the hole-free or Volta PP provides the best phase-shifting properties, which explains its current success and is reflected in numerous applications in cryo-TEM and materials science. 

The simulated images of a monolayer graphene flake show that an application of PPs in high-resolution TEM studies of materials or nano science samples is promising. The simulations show that PPs can improve the interpretability of high-resolution TEM images.

The importance of several parameters to obtain a realistic image appearance is studied. Especially the consideration of a finite size of the zero-order beam of unscattered electrons in the plane of the PP is necessary to yield not only a better quantitative description of the achievable contrast and visibility but also a correct appearance of artifacts such as fringe artifacts at object borders. 

\section*{Acknowledgements}

S.H. and R.A. acknowledge funding by German Research Foundation (DFG project He 7675/1-1) and by the Spanish MICINN (PID2019-104739GB-100/AEI/10.13039/501100011033) and from the European Union H2020 program “ESTEEM3” (Grant number 823717) European Union H2020 ``Graphene Flagship'' CORE 3 (Grant number  881603).

\vspace{100px}

\section*{Supplementary information}        

        \setcounter{table}{0}
        \renewcommand{\thetable}{S\arabic{table}}%
        \setcounter{figure}{0}
        \renewcommand{\thefigure}{S\arabic{figure}}%
        \setcounter{section}{0}
        \renewcommand{\thesection}{SI\arabic{section}}%

\section{Residual aberrations}
\label{SIS1}
Residual aberrations were set to 0 for the calculation of aC NPs and bacteriophages. For the simulation of the graphene lattice, the parameters were set to the values as detailed in Table~\ref{ST:ResABs}.

\begin{table}[ht]
    \centering
    \begin{tabular}{c|c|c}
 & Magnitude & angle/º \\
 \hline
 Spherical aberration C\textsubscript{S} ($C_3$) & 1 $\upmu$m & \\
     Twofold astigmatism $a_1$ & 0 nm & 0 \\
     Threefold astigmatism $a_2$ & 20 nm & 45 \\
     Fourfold astigmatism $a_3$ & 300 nm & 16 \\
     Fivefold astigmatism $a_4$ & 20 $\upmu$m & 124 \\
     Axial coma $b_2$ & 5~nm & 96 \\
     Axiat star $s_3$ & 300 nm & 55 \\
    \end{tabular}
    \caption{Residual aberrations considered for the image simulation of the graphene sample.}
    \label{ST:ResABs}
\end{table}


\newpage

\section{Additional images of T4 bacteriophages}
\label{SIS2}
\subsection*{Conventional TEM and Ideal PP TEM}

\begin{figure}[ht]
    \centering
    \includegraphics[width=1\linewidth]{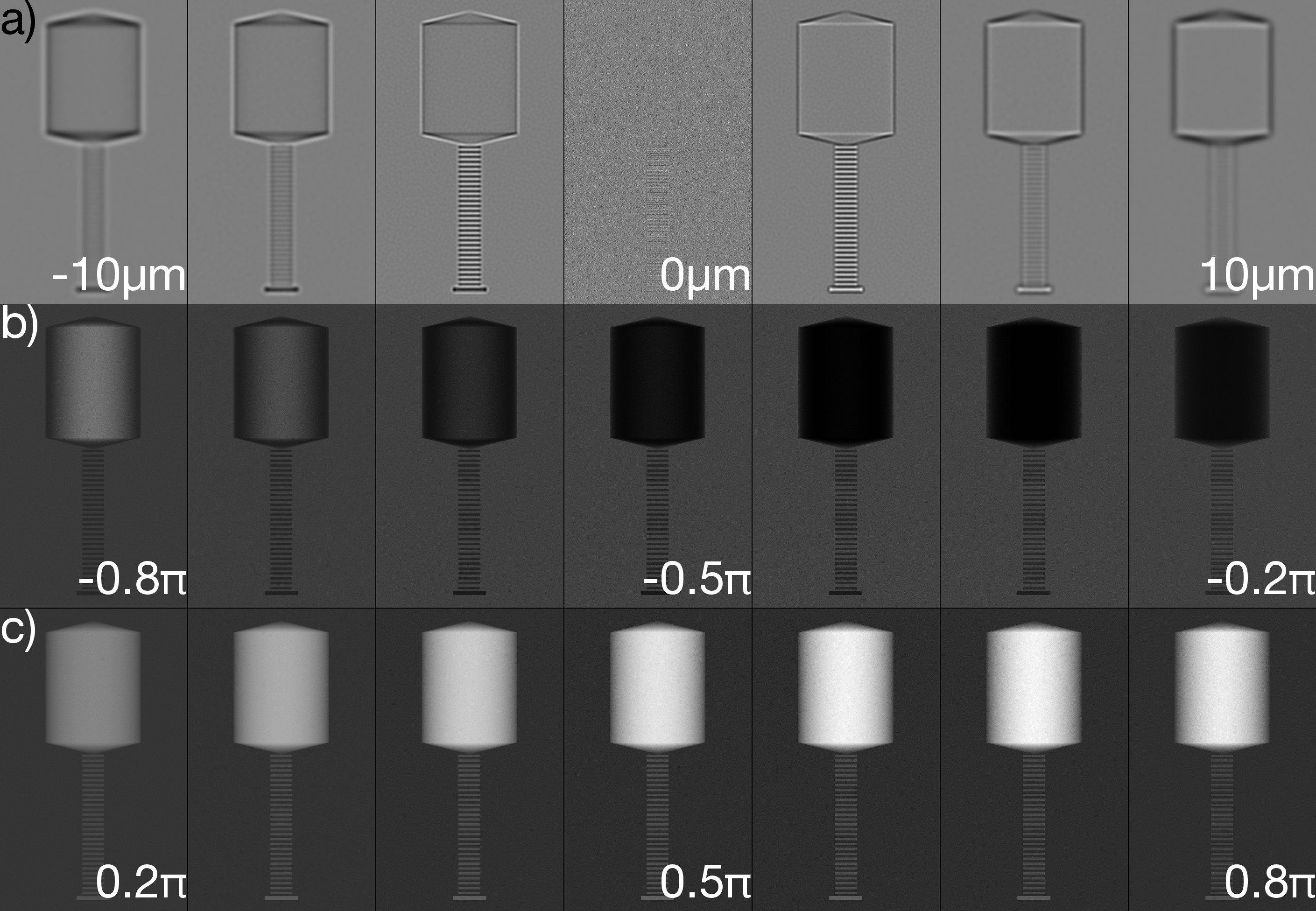}
    \caption{Simulated images of a T4 bacteriophage for: (a) conventional TEM for $Z$ with linear steps between -10 and 10 $\upmu$m. At a moderate defocus, the bacteriophage is well distinguishable and image distortions are minor. For higher $Z$, the bacteriophage appears blurred and the tail repeat is no longer distinguishable. (b,c) Ideal PP TEM for (b) negative applied phase shift and (c) positive applied phase shift. The phase shift series reveal that while the tail shows maximum contrast at $\pm$0.5$\uppi$ applied phase shift, the body, being a strong phase object, reaches its maximum contrast at -0.4$\uppi$ and 0.6$\uppi$. The contrast is adjusted separately for (a) and (b,c) to visualize the effects observed in CTEM. Width of a single image is 152~nm. }
    \label{SF:T4_CtemITEM}
\end{figure}

\newpage

\subsection*{Zernike and Hilbert PP TEM}

\begin{figure}[ht]
    \centering
    \includegraphics[width=0.5\linewidth]{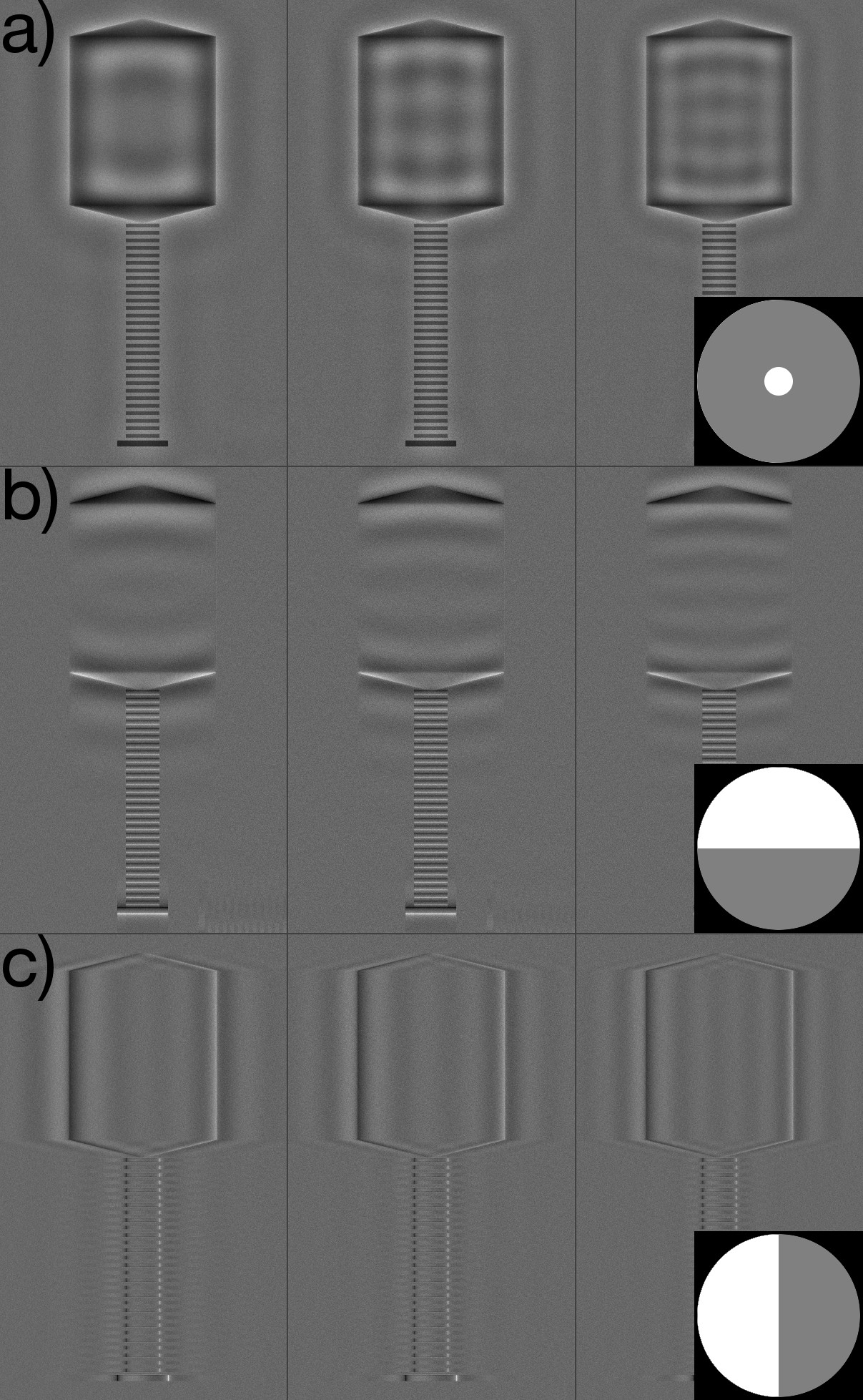}
    \caption{Simulated images of a T4 bacteriophage for: (a) the Zernike PP and the Hilbert PP with (b) horizontal orientation and (c) vertical orientation with respect to the phage for three different radii $r_{ZPP}$ and distances $d_{HPP}$~=~350 (1\textsuperscript{st} column), 425 (2\textsuperscript{nd}) and 500 nm (3\textsuperscript{rd}), respectively. While the contrast of the tail is barely affected by the variation, the weak PC visible in the body for (a) $r_{ZPP}$~=~350~nm vanishes almost completely for higher values. The intensity of the introduced fringing artifacts decreases with increasing radius (distance) while its periodicity increases. The contrast is adjusted separately for (a) and (b,c). Width of a single image is 152~nm. }
    \label{SF:T4_ZtemHilb}
\end{figure}

\newpage

\subsection*{HFPP TEM}

\begin{figure}[ht]
    \centering
    \includegraphics[width=0.75\linewidth]{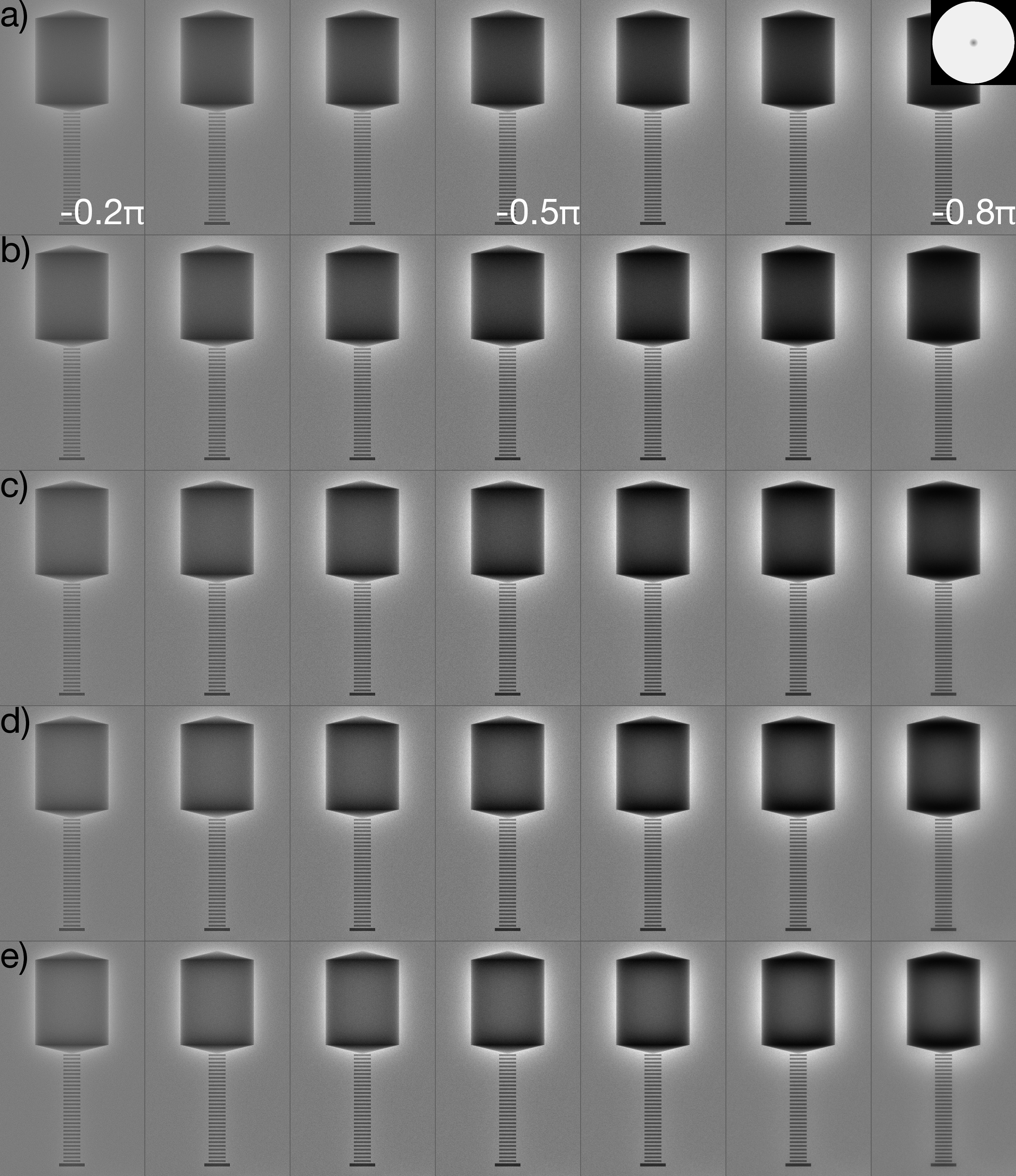}
    \caption{Simulated images of a T4 bacteriophage for the HFPP with different $d_{HFPP}$~=(a) 100~nm, (b) 150~nm, (c) 200~nm, (d) 250 and (e) 300~nm at different applied phase shifts of -0.2$\uppi$ (1\textsuperscript{st} column) to -0.8$\uppi$  (last column). The results show that the contrast of the phage body decreases with increasing $d_{HFPP}$ for the identical applied phase shift (going down a column) and the maximum contrast is reached at increasingly negative applied phase shifts, being outside of the investigated range. The halo around the phage body as well intensifies with increasing $d_{HFPP}$. The tail contrast reaches its maximum at (a) -0.7$\uppi$ for the smallest $d_{HFPP}$ due to an effective decrease of the applied phase shift due to the employed finite ZOB size of 50~nm. For larger $d_{HFPP}$, the maximum contrast is reached at -0.5$\uppi$ as expected for a WPO.  }
    \label{SF:T4_HFPP}
\end{figure}

\newpage

\subsection*{Boersch PP TEM}

\begin{figure}[ht]
    \centering
    \includegraphics[width=1\linewidth]{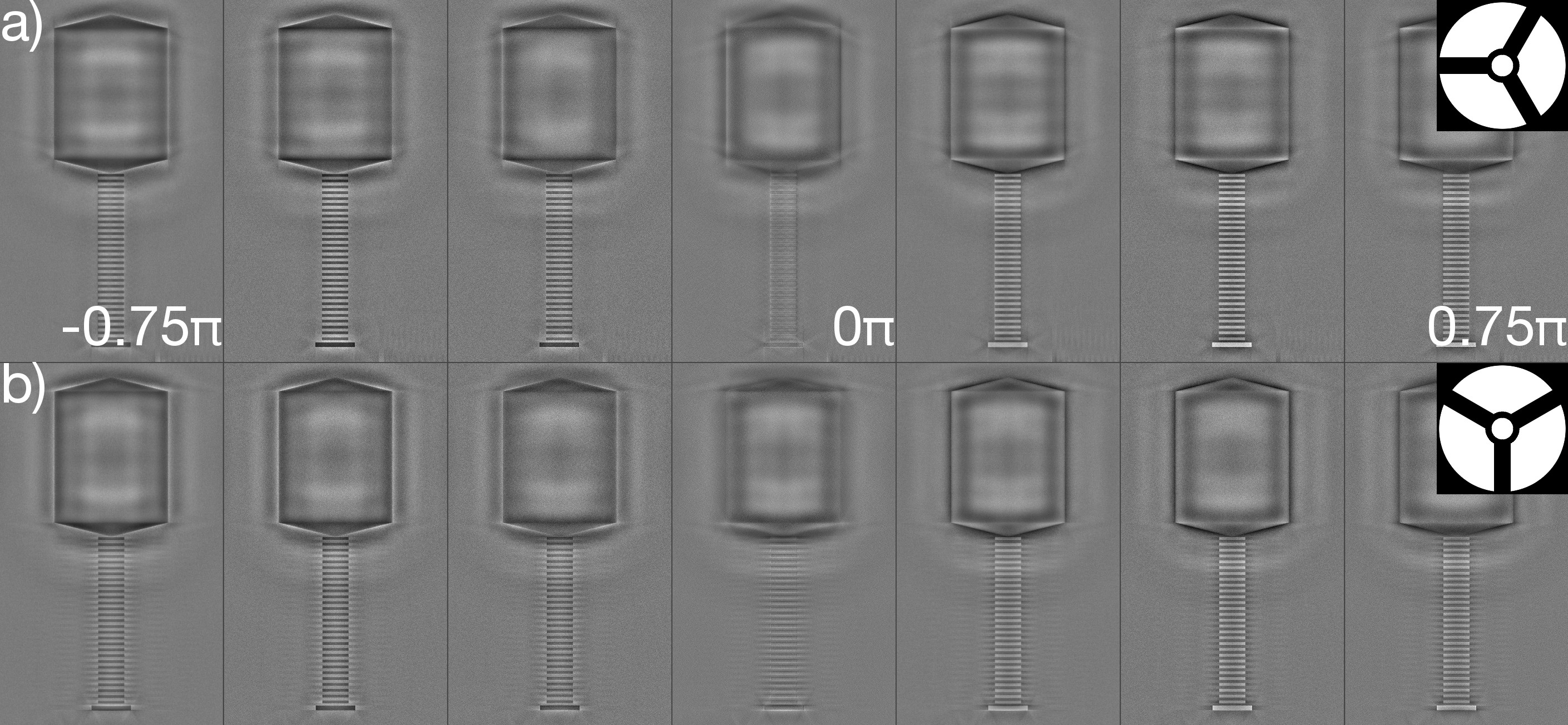}
    \caption{Simulated images of a T4 bacteriophage for the Boersch PP with the bacteriophage aligned (a) horizontally and (b) vertically with respect to the PP for different applied phase shifts between -0.75$\uppi$ and 0.75$\uppi$. The fringing contrast dominates the body and only the triangular edges of the body show a considerable contrast change with the applied phase shift. The contrast merely induced by obstruction of the supporting structure of the BPP is visible in the central column at an applied phase shift of 0$\uppi$. In contrast to the phage body, the tail contrast clearly is inverted in contrast from positive PC for a negative applied phase shift to negative PC for a positive phase shift for both phage orientations.  }
    \label{SF:T4_Boersch}
\end{figure}

\newpage

\subsection*{Zach PP TEM}

\begin{figure}[ht]
    \centering
    \includegraphics[width=0.95\linewidth]{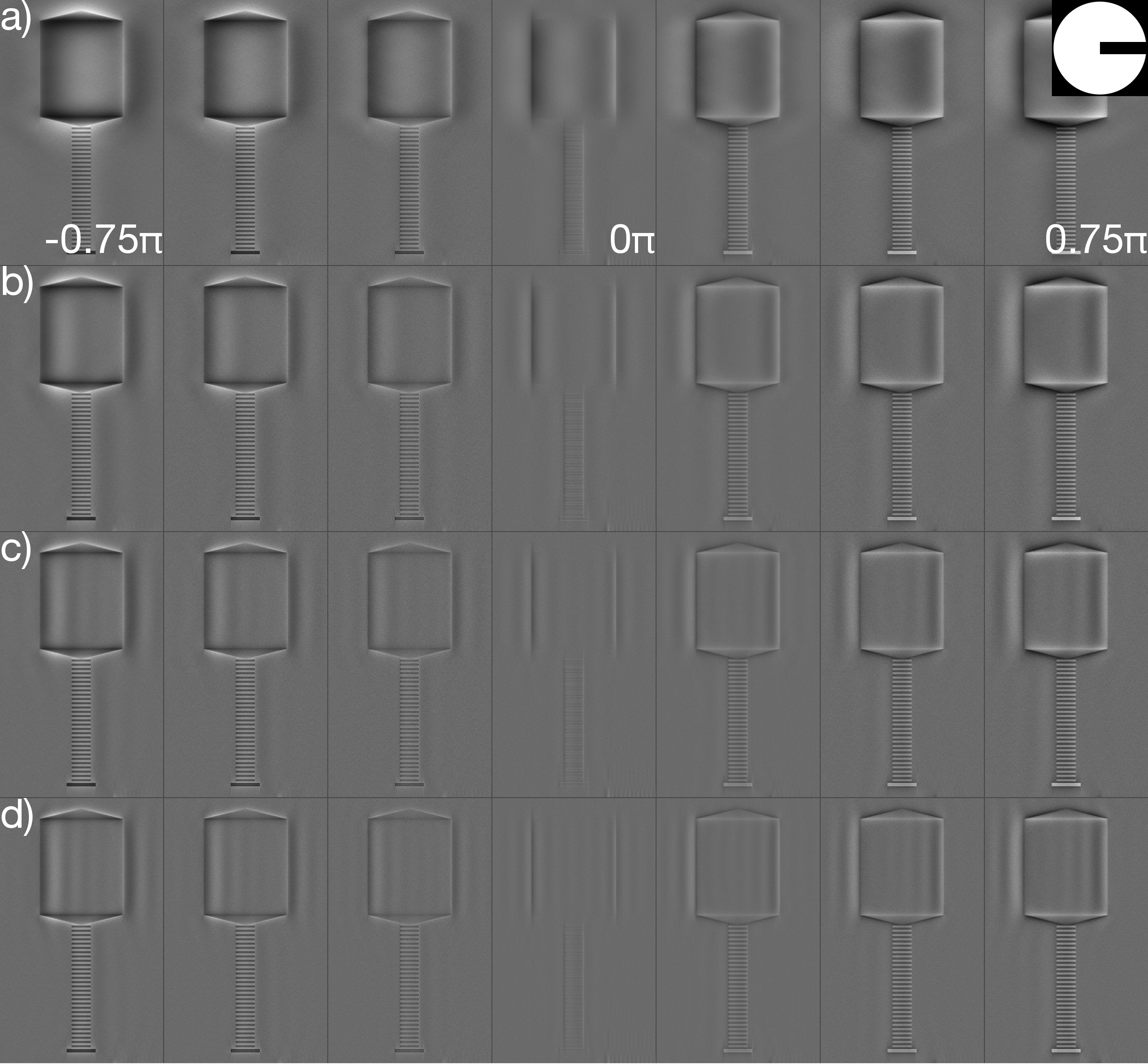}
    \caption{Simulated images of a horizontally aligned T4 bacteriophage for the Zach PP with different applied phase shifts between -0.75$\uppi$ and 0.75$\uppi$ and different distances to the ZOB $d_{Zach}$~= (a) 500~nm, (b) 1000~nm, (c) 1500~nm and (d) 2000~nm. The contrast induced by the obstruction of electrons by the Zach rod is visible in the central column and decreases strongly with increasing $d_{Zach}$. The application of a phase shift induces a clear change in PC in the tail and a varying contrast throughout the body, which decreases with increasing $d_{Zach}$. In additon, fringe contrast appears for larger $d_{Zach}$ within the body.}
    \label{SF:T4_Zach}
\end{figure}

\newpage

\subsection*{Laser PP TEM}

\begin{figure}[ht]
    \centering
    \includegraphics[width=1\linewidth]{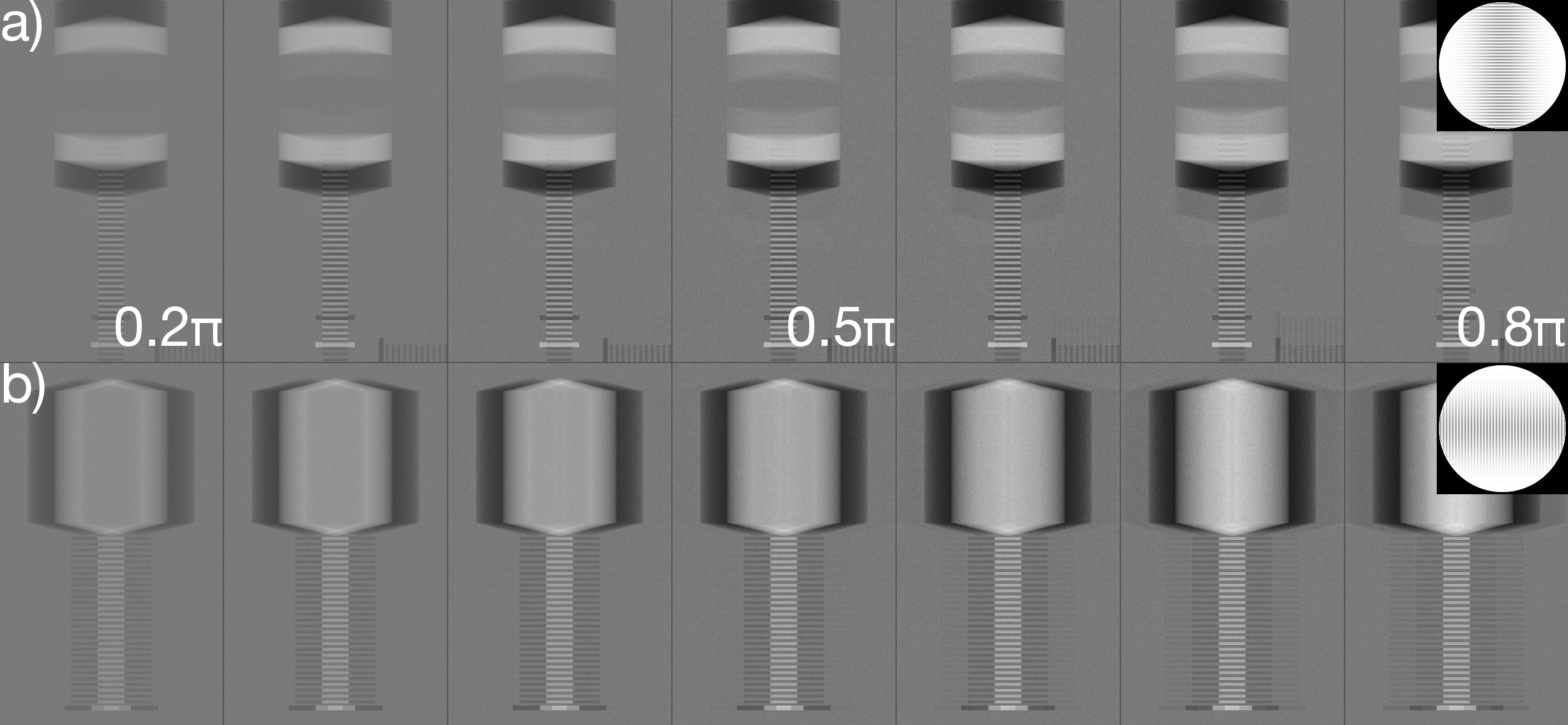}
    \caption{Simulated images of (a) the horizontal and (b) the vertical T4 bacteriophage with the Laser PP at different applied phase shifts from 0.2$\uppi$ to 0.8$\uppi$. The appearance of the phage body is dominated by the presence of shadow images, which overlap with the body. The contrast is increasing with increasing applied phase shift. The tail in (a) is overlayed by a slowly varying contrast change caused by the shadow images but appears homogeneous in (b) as the shadow images do not overlap with the tail.}
    \label{SF:T4_Laser}
\end{figure}

\newpage

\subsection*{Noisy convetional TEM and Ideal PP TEM with reduced electron dose}

\begin{figure}[ht]
    \centering
    \includegraphics[width=0.8\linewidth]{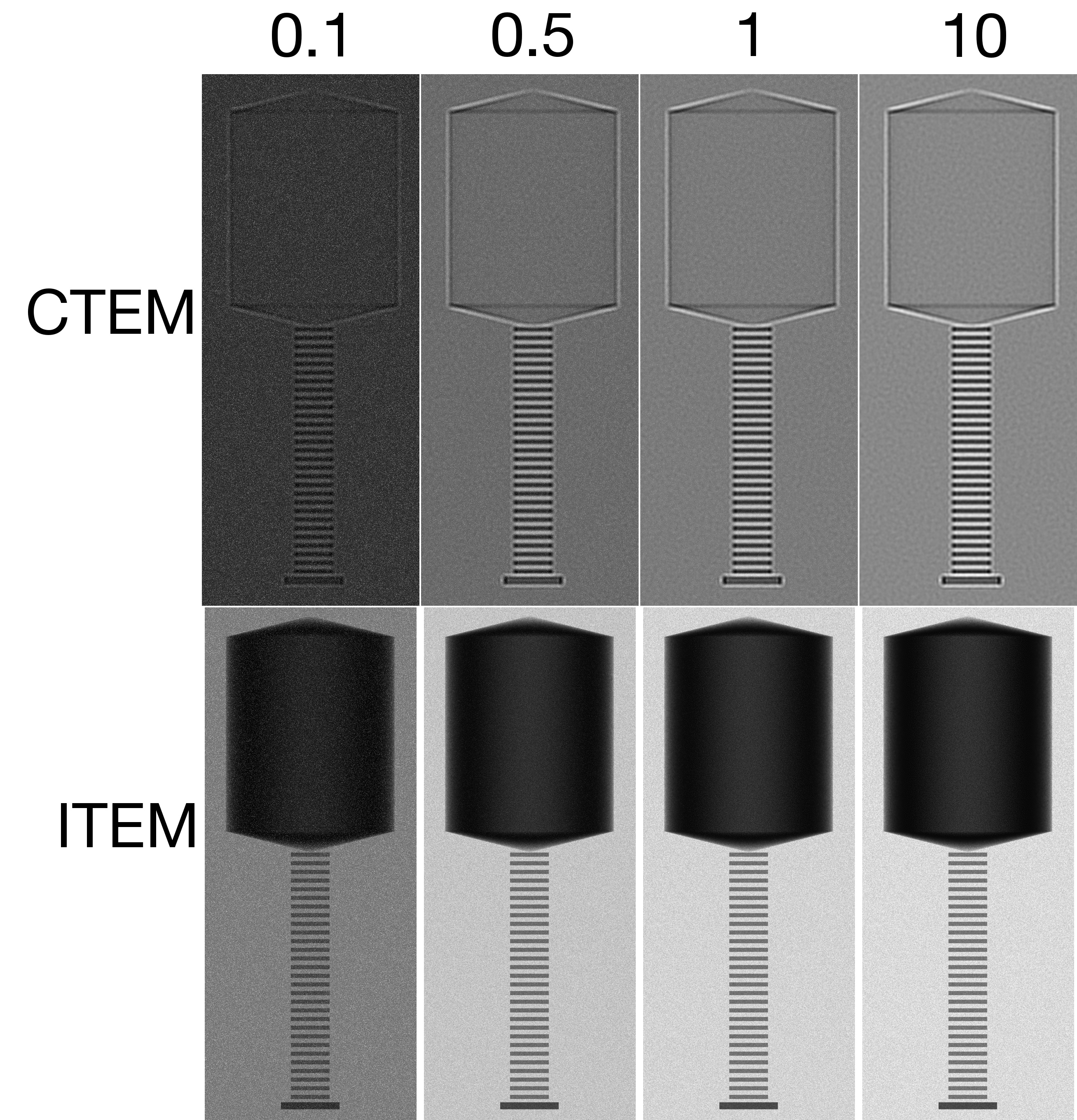}
    \caption{Simulated CTEM (1\textsuperscript{st} row) at $Z$~=~-2~$\upmu$m and ideal PP TEM (2\textsuperscript{nd} row) images of the vertical T4 bacteriophage with varying doses between 0.1 and 10~e\textsuperscript{-1}/A\textsuperscript{2}. Image width of individual images is 100~nm.}
    \label{SF:T4_Noise}
\end{figure}

\newpage

\section{CTEM analysis of aC nanoparticles}
\label{SIS3}

\begin{figure}[ht]
    \centering
    \includegraphics[width=0.9\linewidth]{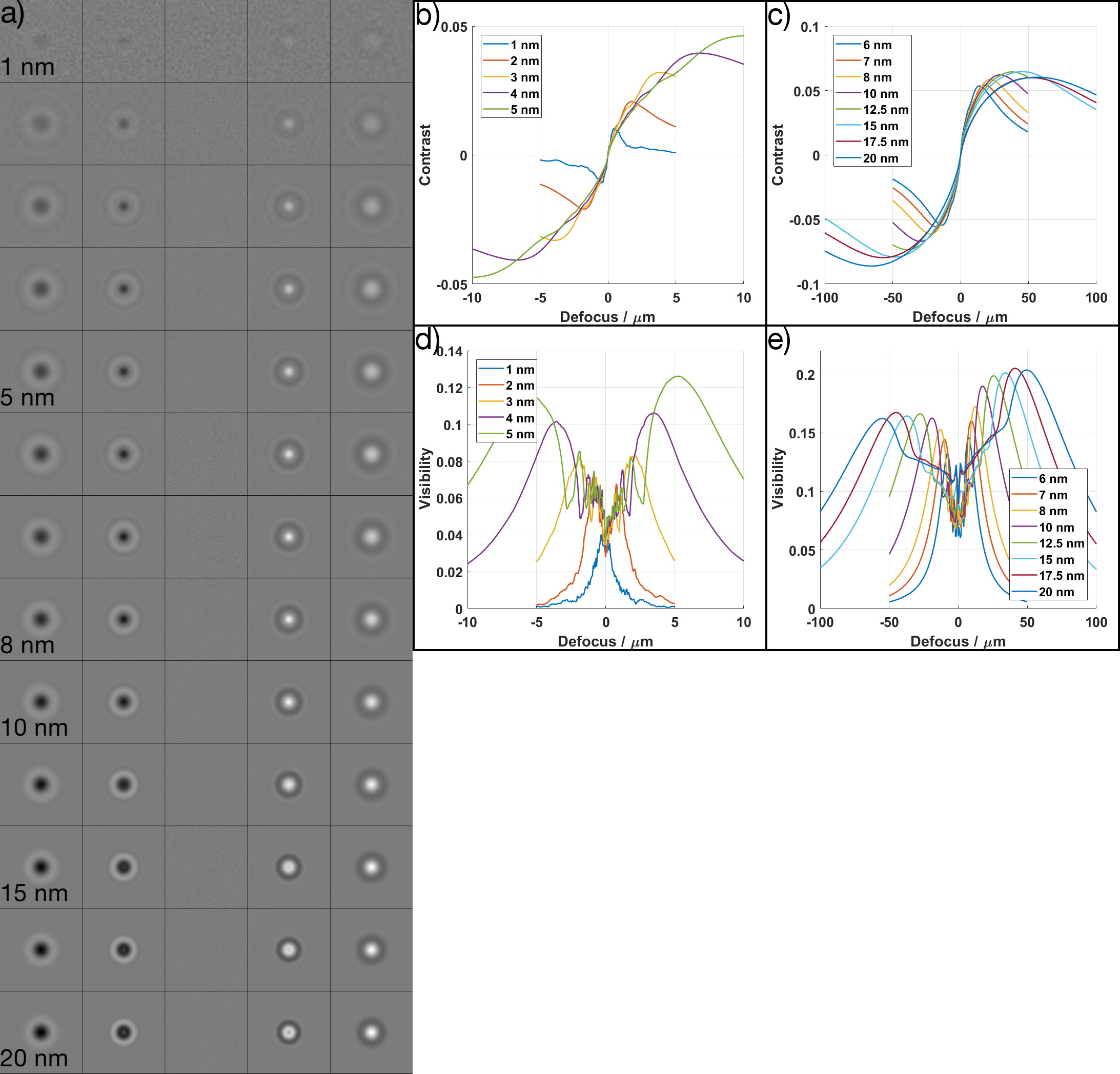}
    \caption{Analysis of CTEM images of amorphous carbon (aC) nanoparticles embedded in ice. (a) Compilation of images of aC nanoparticles with different diameter from 1~nm to 20~nm and for different $Z$~$<$~0 (1\textsuperscript{st} and 2\textsuperscript{nd} column), =~0 (3\textsuperscript{rd} column) and $>$~0 (4\textsuperscript{th} and 5\textsuperscript{th}). The defocus values are listed in Table~\ref{ST:1}. Contrast scales from 0.68 (black) to 0.97 (white). (b,c) Contrast of the NPs in dependence of the applied defocus for diameters between (b) 1 and 5~nm and (c) 6 and 20~nm. The contrast reaches its extrema at increasing $|Z|$. (d,e) Visibility of the NPs in dependence of the applied defocus for diameters between (d) 1 and 5~nm and (e) 6 and 20~nm. It is noted that maximum visibility is reached before maximum contrast.}
    \label{SF:SpCTEM}
\end{figure}

For an in-depth comparison of PP images with conventional (C)TEM  images, simulations of aC NPs in 30~nm of vitrified ice have been performed using the same OEWFs as for the PP simulations and the microscope settings as detailed in Table~1. The studied range of $Z$ was such as to include the maximum reachable contrast. 
Figure~\ref{SF:SpCTEM}a shows the resulting images for the defocus values reaching maximum positive and negative PC (Table~\ref{ST:1}, for half of these values and in focus. The NPs are invisible at $Z$~=~0 as expected for a weak-phase object, which also shows that the effect of the spherical aberration ($C_S$~=~5~mm) can be neglected when regarding the contrast of the NPs. The contrast in the images is a combination of actual phase contrast and fringing introduced by the large defocus. The phase contrast is rather strong for the smallest and highest diameters, which for the small NPs is caused by an increase in contrast transfer and for the larger NPs due to the increase of the object-induced phase shift. At intermediate sizes, the fringing contrast dominates the images.


The contrast and visibility analysis is displayed in Figure~\ref{SF:SpCTEM}b-e. The maximum reachable contrast increases strongly with the NP size before reaching a maximum at a diameter of 15~nm. It is noted that the maximum visibility is reached at smaller values of $Z$ in comparison to the contrast. A strong asymmetry with the sign of $Z$ is observed for the largest NPs. The increased visibility for positive $Z$ is due to the positive sign of the object-induced phase shift and the associated higher contrast/visibility reachable in overfocus, which is as well observed for the negative PC in PP images.

\begin{table}[ht]
    \centering
    \begin{tabular}{c|c|c}
         aC NP diameter & $|Z|_{max}$  \\
    \hline
         1 & 0.45~$\upmu$m  \\
         2 & 1.75~$\upmu$m  \\
         3 & 3.9~$\upmu$m  \\
         4 & 6.8~$\upmu$m  \\
         5 & 10~$\upmu$m \\
         6 & 14~$\upmu$m \\
         7 & 18~$\upmu$m  \\
         8 & 21.5~$\upmu$m  \\
         10 & 29~$\upmu$m  \\
         12.5 & 38~$\upmu$m  \\
         15 & 45~$\upmu$m  \\
         17.5 & 52.5~$\upmu$m \\
         20 & 57~$\upmu$m  \\
    \end{tabular}
    \caption{Values used for CTEM simulations of aC NPs embedded in 30~nm of vitrified ice. }
    \label{ST:1}
\end{table}

\newpage
\section{Additional PP images of aC nanoparticles}
\label{SIS4}
\subsection*{Zernike and HFPP TEM}

\begin{figure}[ht]
    \centering
    \includegraphics[width=0.4\linewidth]{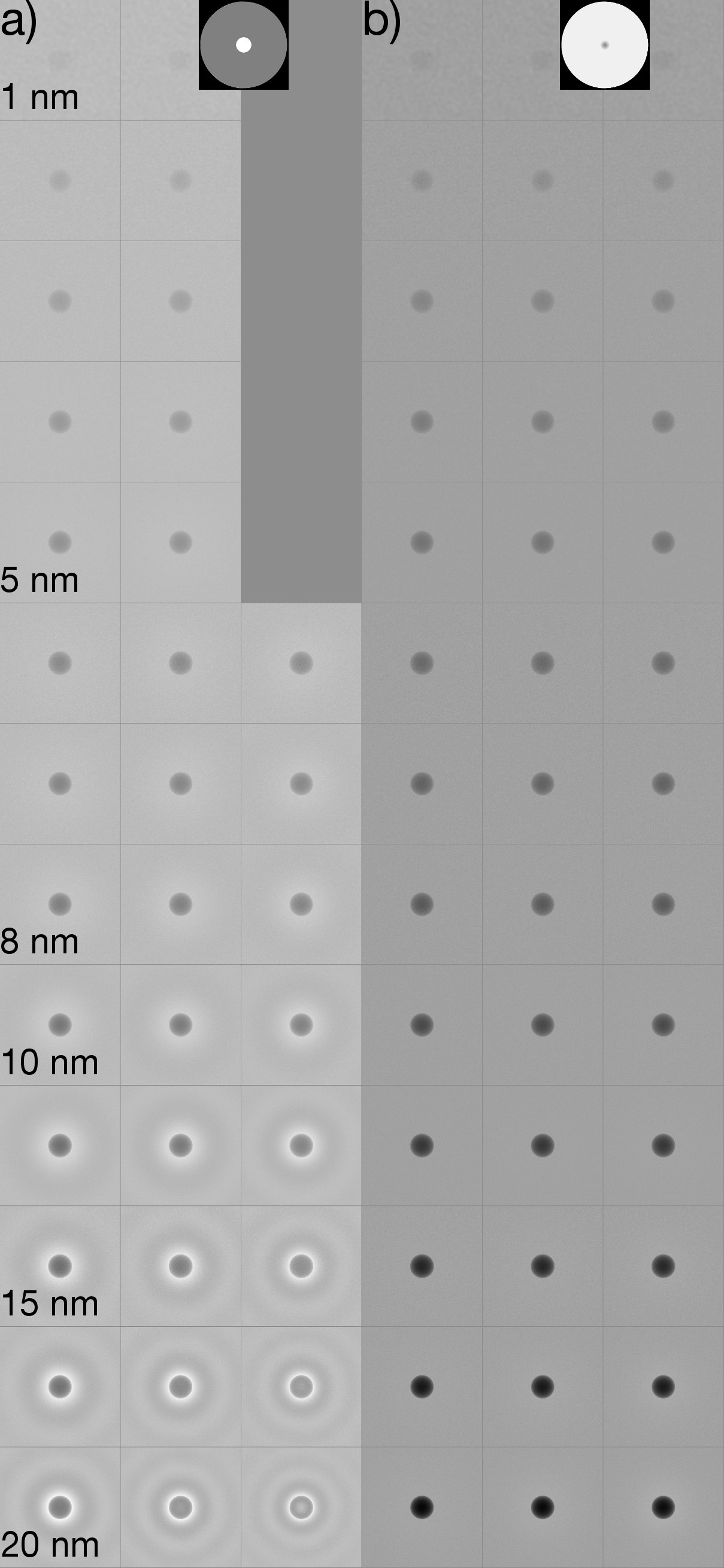}
    \caption{Compilation of images of a spherical aC NP embedded in vitrified ice with diameters between 1 and 20~nm for (a) the Zernike PP and (b) the HFPP for different cut-on frequencies: (a) $r_{ZPP}$~=~350~nm (1\textsuperscript{st} column), 425~nm (2\textsuperscript{nd}) and 500~nm (3\textsuperscript{rd}); (b) $d_{HFPP}$~=~100~nm (1\textsuperscript{st} column), 150~nm (2\textsuperscript{nd} and 200~nm 3\textsuperscript{rd} column). (a) The decreasing PC with increasing  $r_{ZPP}$ for the larger NPs is visible, which also goes in hand with an increase of the fringing artifact at increased periodicity. Due to low sampling of the BFP, the 2\textsuperscript{nd} and 3\textsuperscript{rd} value for $r_{ZPP}$ lead to the identical number of pixels. (b) While the NPs appear without noticeable artifacts for the smallest $d_{HFPP}$, the halo around the large NPs is gaining importance with increasing $d_{HFPP}$.}
    \label{SF:Sp_ZernHFPP}
\end{figure}

\newpage

\subsection*{Boersch PP TEM}

\begin{figure}[ht]
    \centering
    \includegraphics[width=0.47\linewidth]{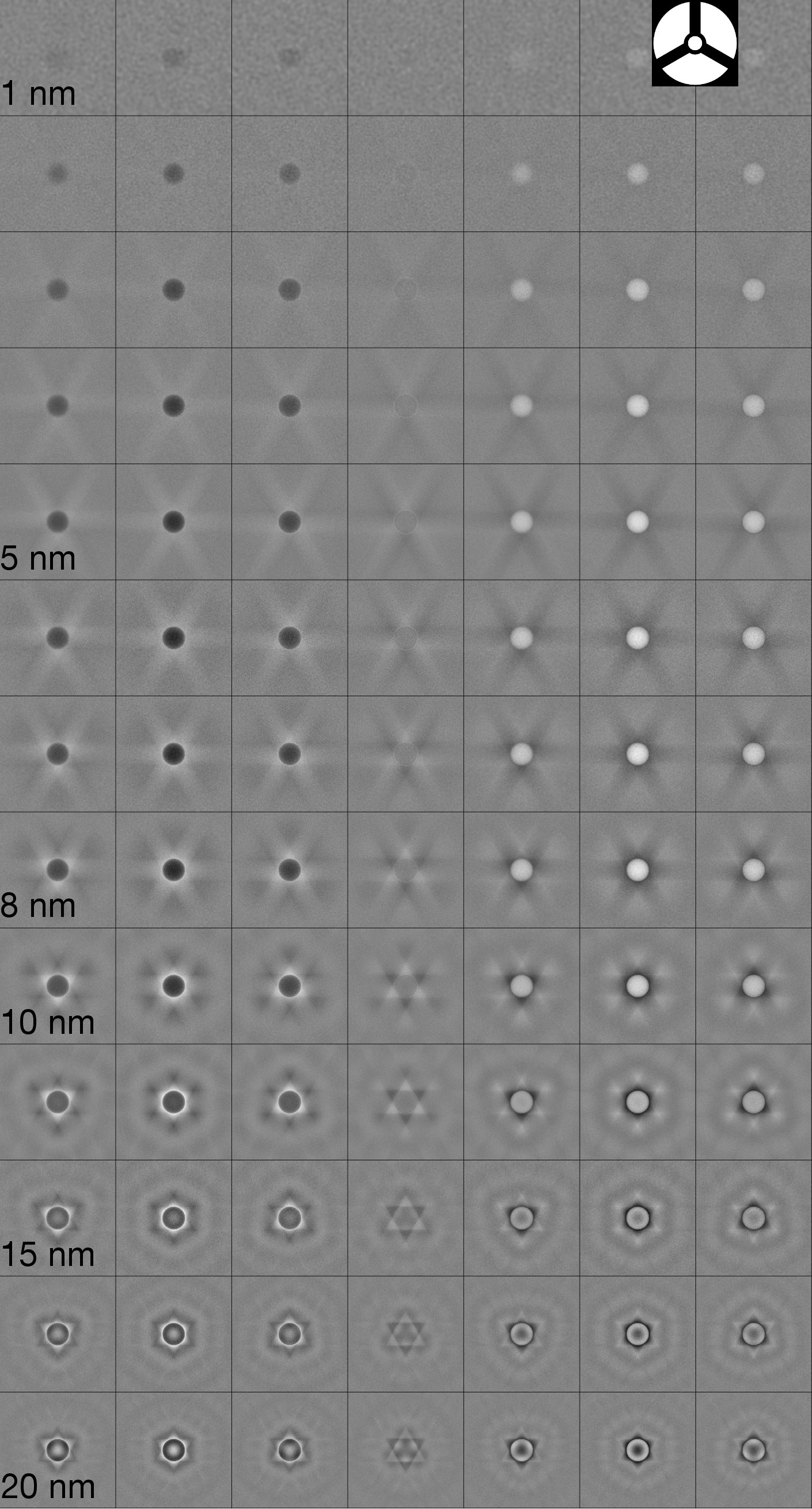}
    \caption{Compilation of images of a spherical aC NP embedded in vitrified ice with diameters between 1 and 20~nm for the BPP at varying induced phase shift between -0.75$\uppi$ (1\textsuperscript{st} column), 0$\uppi$ (4\textsuperscript{th} column) and 0.75$\uppi$ (7\textsuperscript{th} column). The changing PC is well observable until a NP diameter of 8~nm before fringing artifacts start to dominate as well the interior of the NPs.}
    \label{SF:Sp_Boersch}
\end{figure}

\newpage

\subsection*{Zach PP TEM at 500 nm distance to the ZOB}

\begin{figure}[ht]
    \centering
    \includegraphics[width=0.47\linewidth]{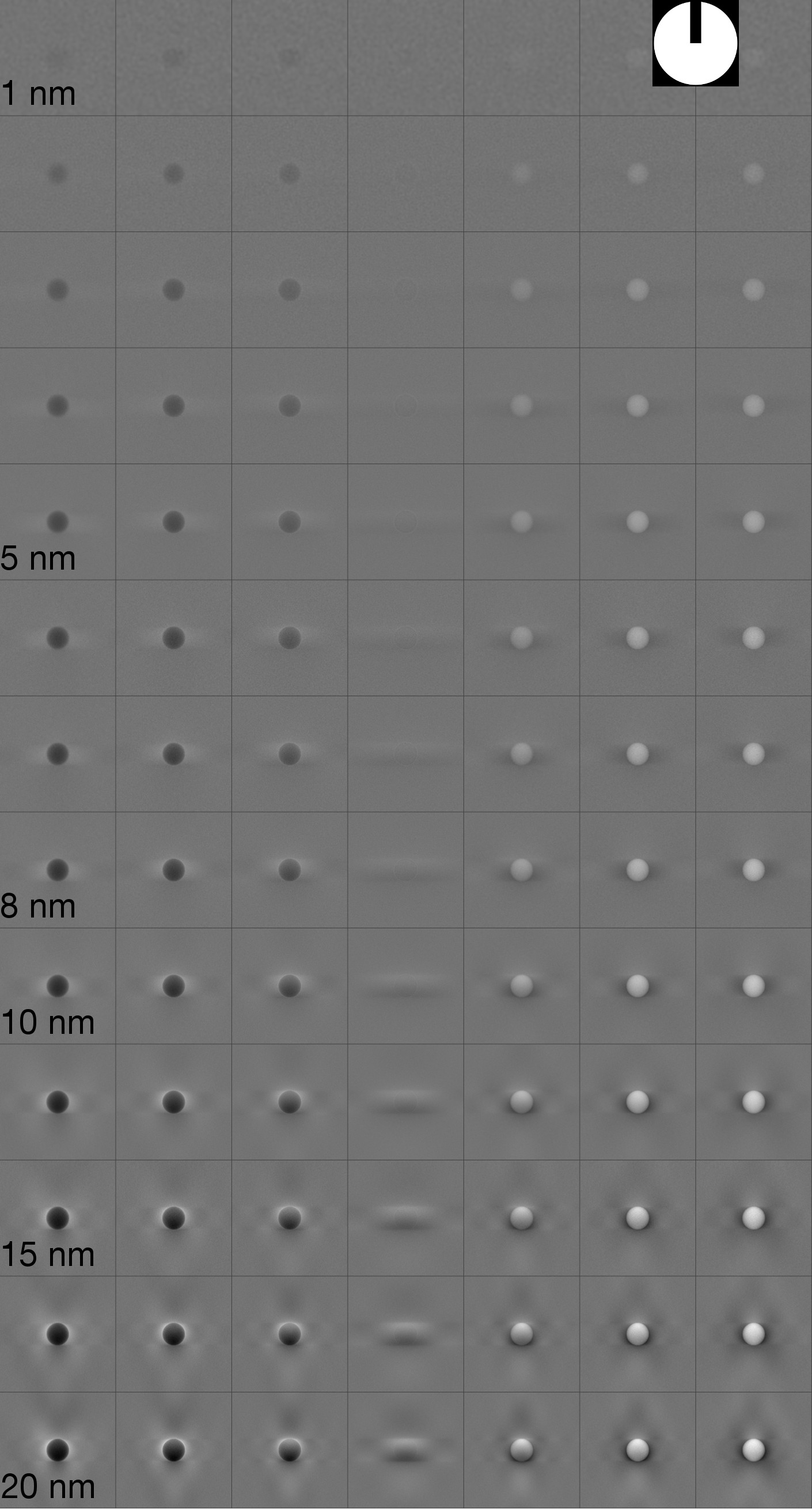}
    \caption{Compilation of images of a spherical aC NP embedded in vitrified ice with diameters between 1 and 20~nm for the Zach PP positioned at a distance of $d_{Zach}$~=~500~nm at varying induced phase shift between -0.75$\uppi$ (1\textsuperscript{st} column), 0$\uppi$ (4\textsuperscript{th} column) and 0.75$\uppi$ (7\textsuperscript{th} column). The PC of the NPs inverts for all NP diameters but the inhomogeneous halo around the NP increases in intensity with NP diameter.}
    \label{SF:Sp_Zach500}
\end{figure}

\newpage

\subsection*{Zach PP TEM at 2000 nm distance to the ZOB}

\begin{figure}[ht]
    \centering
    \includegraphics[width=0.47\linewidth]{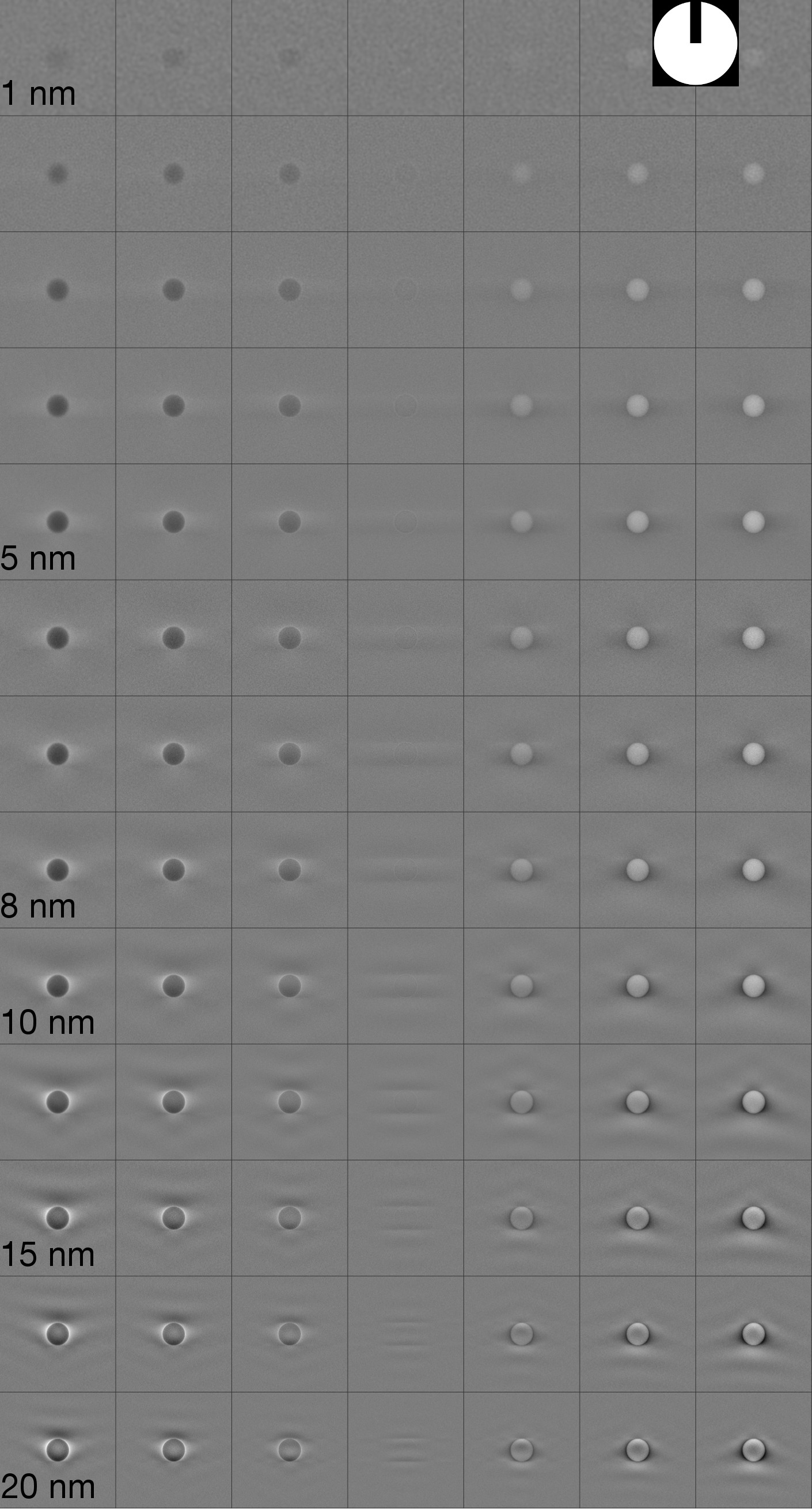}
    \caption{Compilation of images of a spherical aC NP embedded in vitrified ice with diameters between 1 and 20~nm for the Zach PP positioned at a distance of $d_{Zach}$~=~2000~nm at varying induced phase shift between -0.75$\uppi$ (1\textsuperscript{st} column), 0$\uppi$ (4\textsuperscript{th} column) and 0.75$\uppi$ (7\textsuperscript{th} column). In comparison to a shorter distance (Fig.~\ref{SF:Sp_Zach500}), the PC is reduced and the NPs appear with inhomogeneous contrast starting from a NP diameter of 10~nm. }
    \label{SF:Sp_Zach2000}
\end{figure}

\newpage

\subsection*{Tunable Ampere PP TEM}

\begin{figure}[ht]
    \centering
    \includegraphics[width=0.67\linewidth]{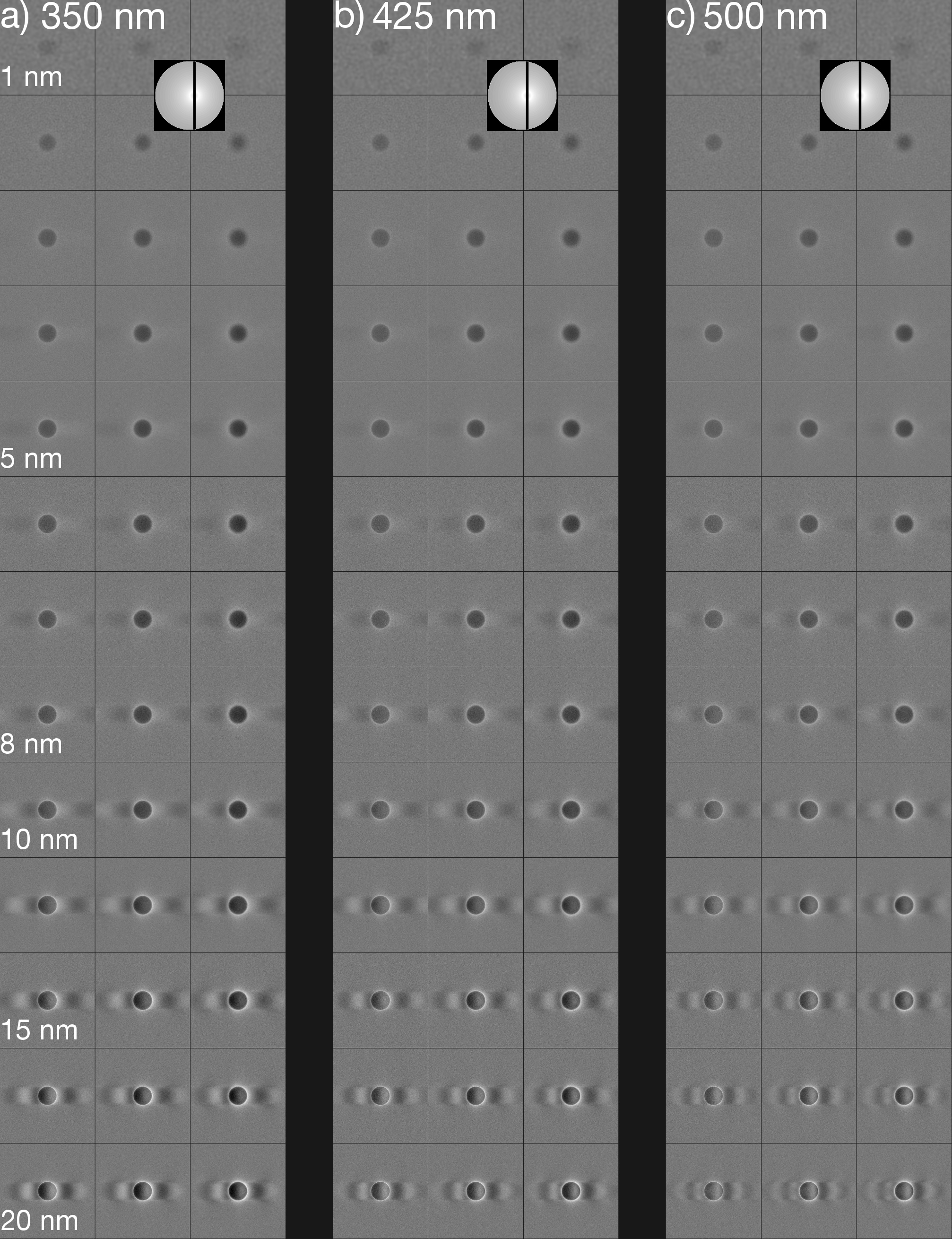}
    \caption{Compilation of images of a spherical aC NP embedded in vitrified ice with diameters between 1 and 20~nm for the TAPP positioned at different distances (a) $d_{TAPP}$~=~350~nm, (b) 425~nm and (c) 500~nm. For each distance, the three columns correspond to an applied current of 2~mA (1\textsuperscript{st} column), 3~mA (2\textsuperscript{nd} column) and 4~mA (3\textsuperscript{rd} column). The PC increases with the applied current and decreases with the distance. The NPs appear with homogeneous contrast surrounded by a bright halo up to a diameter of 5~nm. The halo increases in intensity and turns into fringing artifacts for increasing NP diameter. Starting from a diameter of 10~nm, the NPs appear with inhomogeneous contrast.}
    \label{SF:Sp_TAPP}
\end{figure}

\newpage

\subsection*{Laser PP TEM}

\begin{figure}[ht]
    \centering
    \includegraphics[width=1\linewidth]{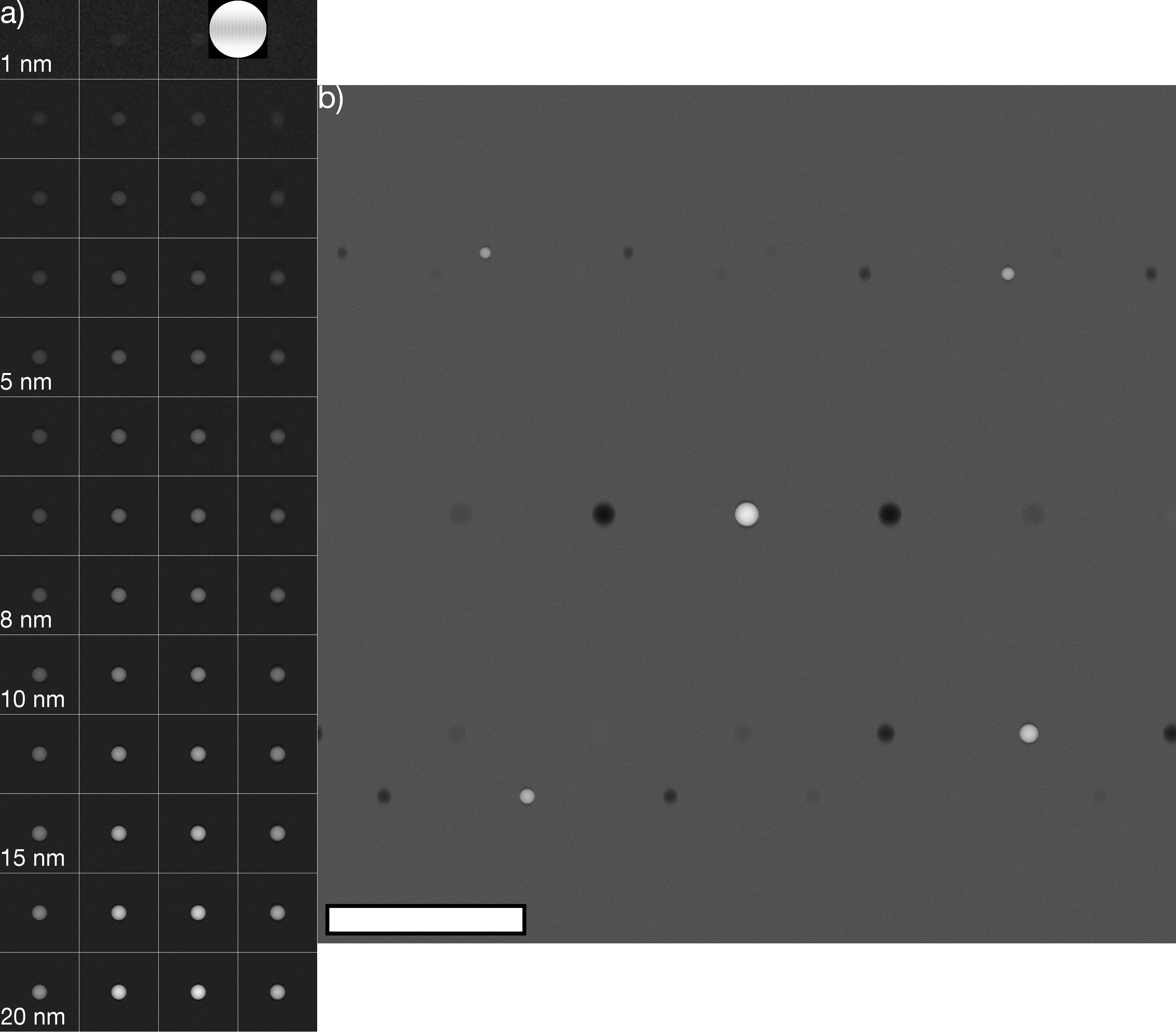}
    \caption{(a) Compilation of images of a spherical aC NP embedded in vitrified ice with diameters between 1 and 20~nm for the Laser PP at different induced phase shifts between 0.25$\uppi$ (1\textsuperscript{st} column) and 1$\uppi$ (4\textsuperscript{th} column). The NPs appear with homogeneous negative PC surrounded by a thin dark halo for all diameters. (b) Several shadow images appear in horizontal orientation in an example image of the NPs with sizes between 5 and 12.5~nm with an applied phase shift of 0.5$\uppi$. }
    \label{SF:Sp_Laser}
\end{figure}

\newpage

\subsection*{Noisy convetional TEM and Ideal PP TEM with reduced electron dose}

\begin{figure}[ht]
    \centering
    \includegraphics[width=\linewidth]{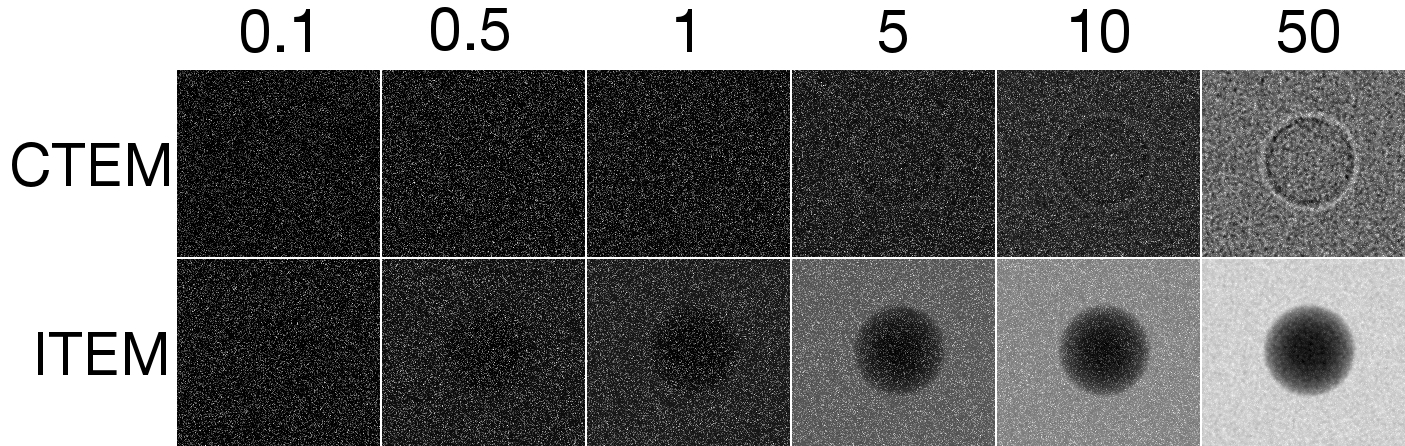}
    \caption{Simulated CTEM (1\textsuperscript{st} row) and ideal PP TEM (2\textsuperscript{nd} row) images of a spherical aC NP with a diameter of 5~nm embedded in 30~nm of vitrified ice at $Z$~=~-70~nm with varying doses between 0.1 and 50~e\textsuperscript{-1}/A\textsuperscript{2}.}
    \label{SF:Sp_Noise}
\end{figure}

\newpage

\section{Additional images of graphene simulations}
\label{SIS5}
\begin{figure}[ht]
    \centering
    \includegraphics[width=0.87\linewidth]{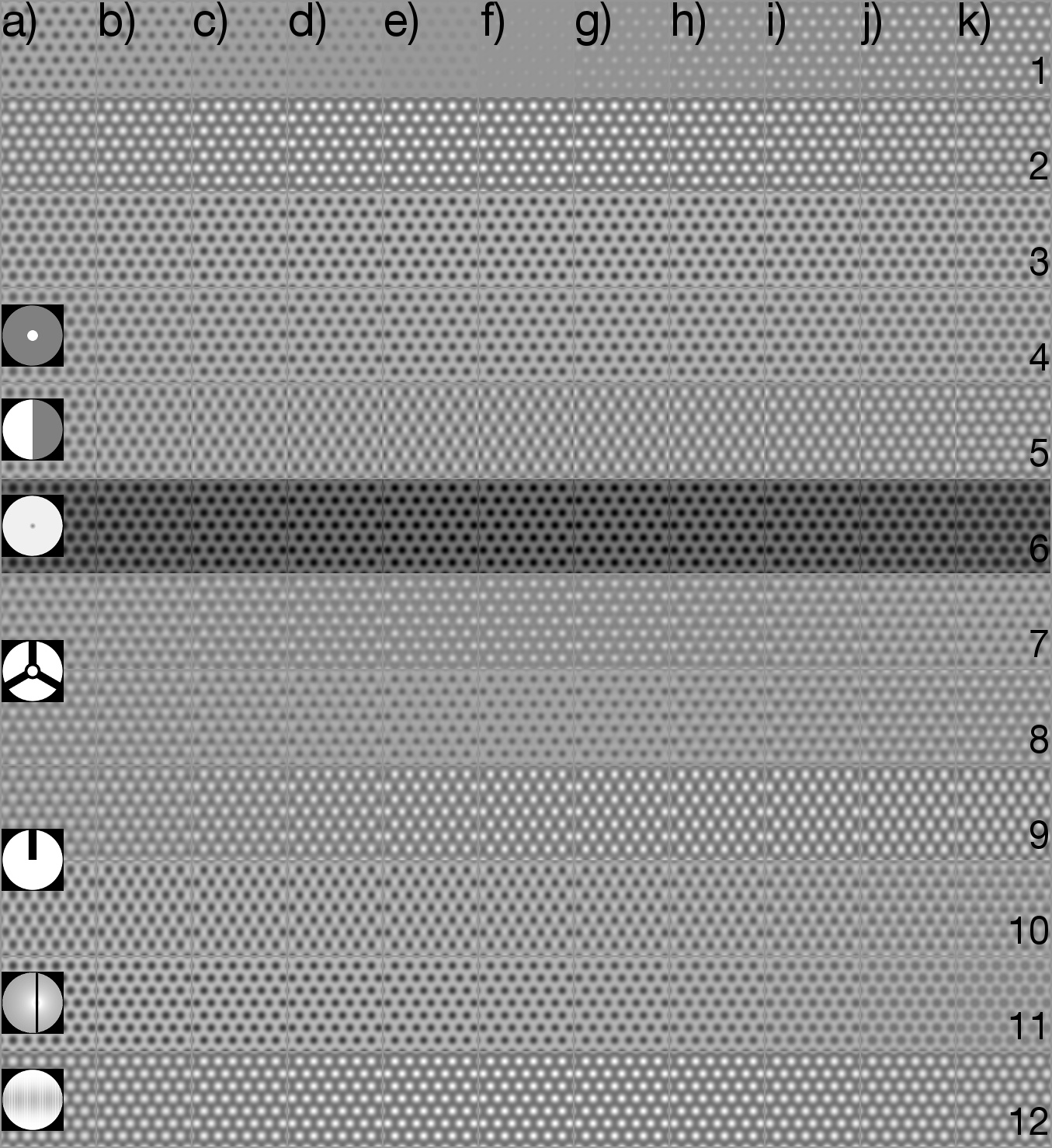}
    \caption{Compilation of images of a graphene monolayer for different $Z$ between (a) -5~nm, (f) 0~nm and (k) +5 nm and different PPs: CTEM (1\textsuperscript{st} row), Ideal PP for $\pm\uppi$/2 (2\textsuperscript{nd} and 3\textsuperscript{rd} row), ZPP (4\textsuperscript{th} row), HPP (5\textsuperscript{th} row), HFPP (6\textsuperscript{th} row), BPP for $\pm\uppi$/2 (7\textsuperscript{th} and 8\textsuperscript{th} row), Zach PP for $\pm\uppi$/2 (9\textsuperscript{th} and 10\textsuperscript{th} row), TAPP at 1.5 mA (11\textsuperscript{th} row) and the LPP (12\textsuperscript{th} row). Width of individual images is 1.6 nm.}
    \label{SF:Graphene}
\end{figure}

\newpage

\section{Image analysis for varying ZOB size}
\label{SIS6}

\begin{figure}[ht]
    \centering
    \includegraphics[width=1\linewidth]{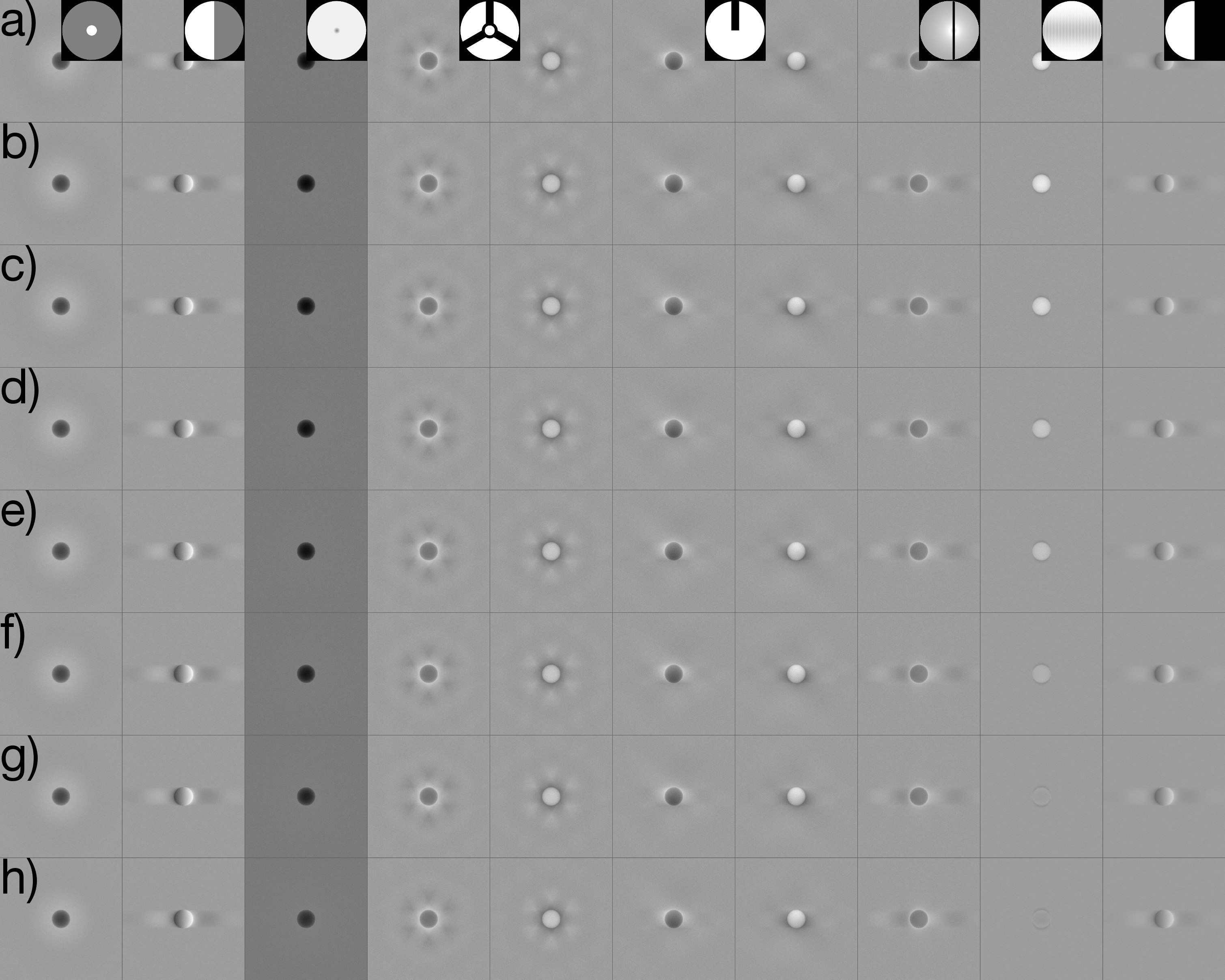}
    \caption{Compilation of images of a spherical aC NP with a diameter of 10~nm embedded in vitrified ice for an increasing diameter of the ZOB in the BFP of (a) 0, (b) 10, (c) 20, (d) 30, (e) 40, (f) 50, (g) 75 and (h) 100~nm.The increasing diameter causes a considerable reduction of fringing and halo artifacts for PPs with sharp edges or obstructing elements. In case of the HFPP and LPP, which exhibit narrow phase profiles, a larger ZOB leads to a decrease of the effectively induced phase shift and thus to a reduction of the contrast.}
    \label{SF:IncNS}
\end{figure}

\newpage

\section{Image analysis for misplaced PPs}
\label{SIS7}

\begin{figure}[ht]
    \centering
    \includegraphics[width=1\linewidth]{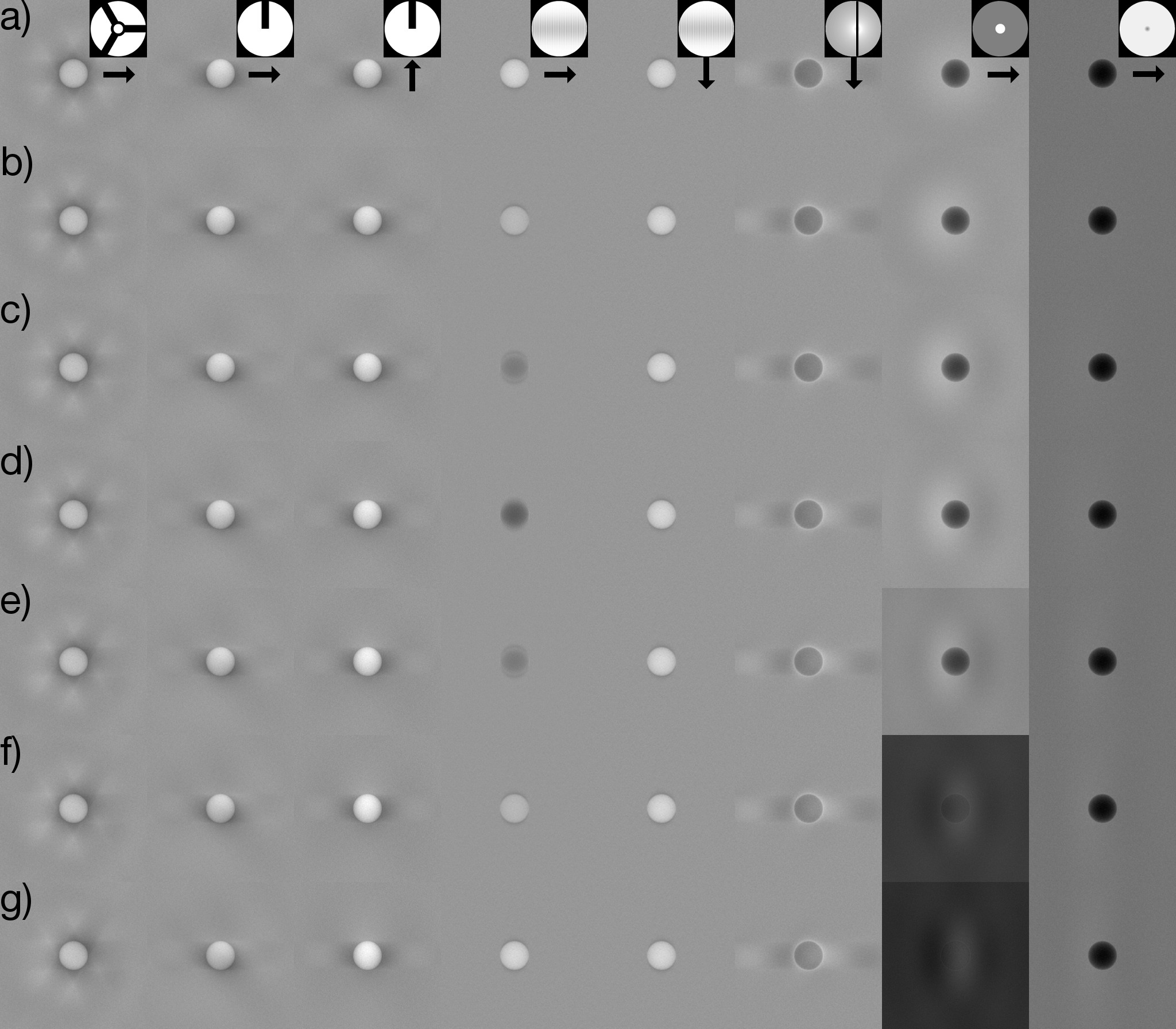}
    \caption{Compilation of images of a 10 nm sized aC NP embedded in vitrified ice for the Boersch PP (1\textsuperscript{st} column), Zach PP (2\textsuperscript{nd} and 3\textsuperscript{rd} column), Laser PP (4\textsuperscript{th} and 5\textsuperscript{th} column), TAPP (6\textsuperscript{th} row), Zernike PP (7\textsuperscript{th} row) and HFPP (8\textsuperscript{th} row) deliberately displaced in the indicated direction. Displacement increases from (a) 0~nm to (g) approximately 500~nm. Exact values are for BPP and ZPP in the MBFP (b-g) 70, 210, 280, 350, 420, 490 nm, for the LPP in the LBFP (b-g) 89, 177, 266, 354, 443, 532 nm and the TAPP, ZPP and HFPP in the regular BFP (b-g) 0, 89, 178, 244, 333, 377, 421 nm. The values differ due to the different employed focal lengths and pixel-based calculation of the displacement.}
    \label{SF:Misplaced}
\end{figure}

Figure~\ref{SF:Misplaced} shows the image appearance of a 10~nm sized aC NP when deliberately misplacing the PP. For the BPP and Zach PP the main effect is a change in the halo around the PP, which is as well observed for the TAPP. The strong change of the phase-shift distribution in horizontal direction for the LPP causes a phase-contrast inversion of the NP upon horizontal displacement of the LPP while no obvious contrast change is observable in vertical direction. The contrast of the NP changes abruptly once the ZOB is not in the ZPP hole anymore. As the phase-shifting profile of the HFPP is induced by the ZOB, a misplacement is considered by an elongated phase-shifting profile, as it would be caused by a drift of the HFPP during the buildup. This drift causes the appearance of a halo around the NP.

\newpage

\section{Image analysis for varying image size}
\label{SIS8}

\begin{figure}[ht]
    \centering
    \includegraphics[width=0.6\linewidth]{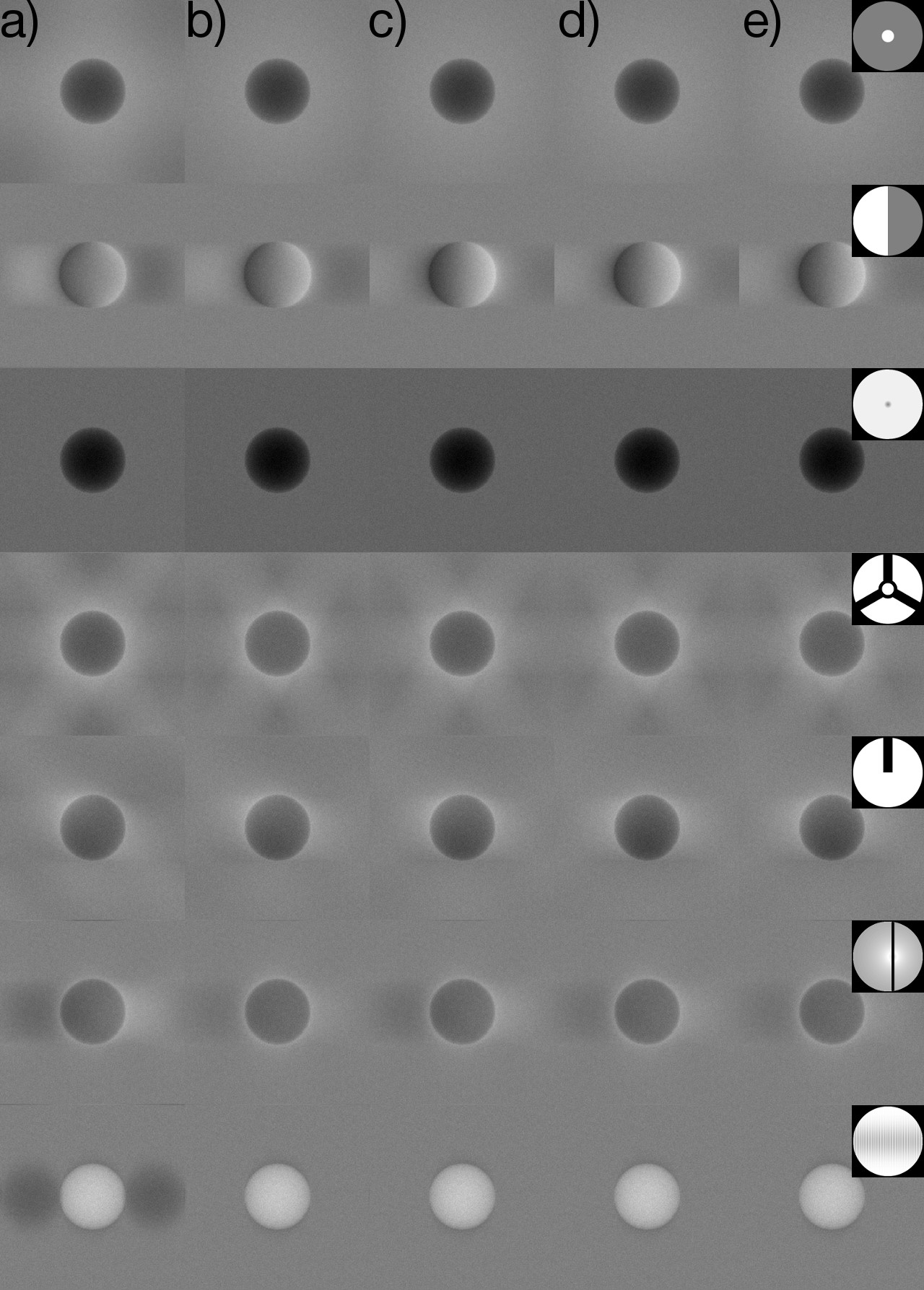}
    \caption{Compilation of images of a 5 nm sized aC NP embedded in vitrified ice for the Zernike PP (1\textsuperscript{st} row), Hilbert PP (2\textsuperscript{nd} row), HFPP (3\textsuperscript{rd} row), Boersch PP (4\textsuperscript{th} row), Zach PP (5\textsuperscript{th} row), TAPP (6\textsuperscript{th} row) and Laser PP (7\textsuperscript{th} row) and increasing total number of pixel of the OEWF of (a) 256, (b) 512, (c) 1024, (d) 2048 and (e) 4096. }
    \label{SF:PixelNumber}
\end{figure}

Figure~\ref{SF:PixelNumber} shows the central part of the simulated images containing a spherical aC NP with diameter of 5~nm embedded in vitrified ice for different employed PPs and total pixel sizes $n_{px}$ of the OEWF. Although the NP itself is nicely sampled already for $n_{px}$~=~256 (Figure~\ref{SF:PixelNumber}a), a considerable contrast change when increasing to $n_{px}$~=~512 (\ref{SF:PixelNumber}b) is observed for almost all PP types. The halo and fringes around the NP appear smoother (e.g. for the ZPP and TAPP), the background intensity in the HFPP image decreases and the shadow images of the Laser PP move to larger distances. The observed changes in the NP appearance are less when further increasing $n_{px}$ and no visible change between $n_{px}$~=~2048 and $n_{px}$~=~4096 (Figure~\ref{SF:PixelNumber}d and e) may be observed.

\newpage

\section{Simulation software}
\label{SIS9}
\begin{figure}[ht]
    \centering
    \includegraphics[width=1\linewidth]{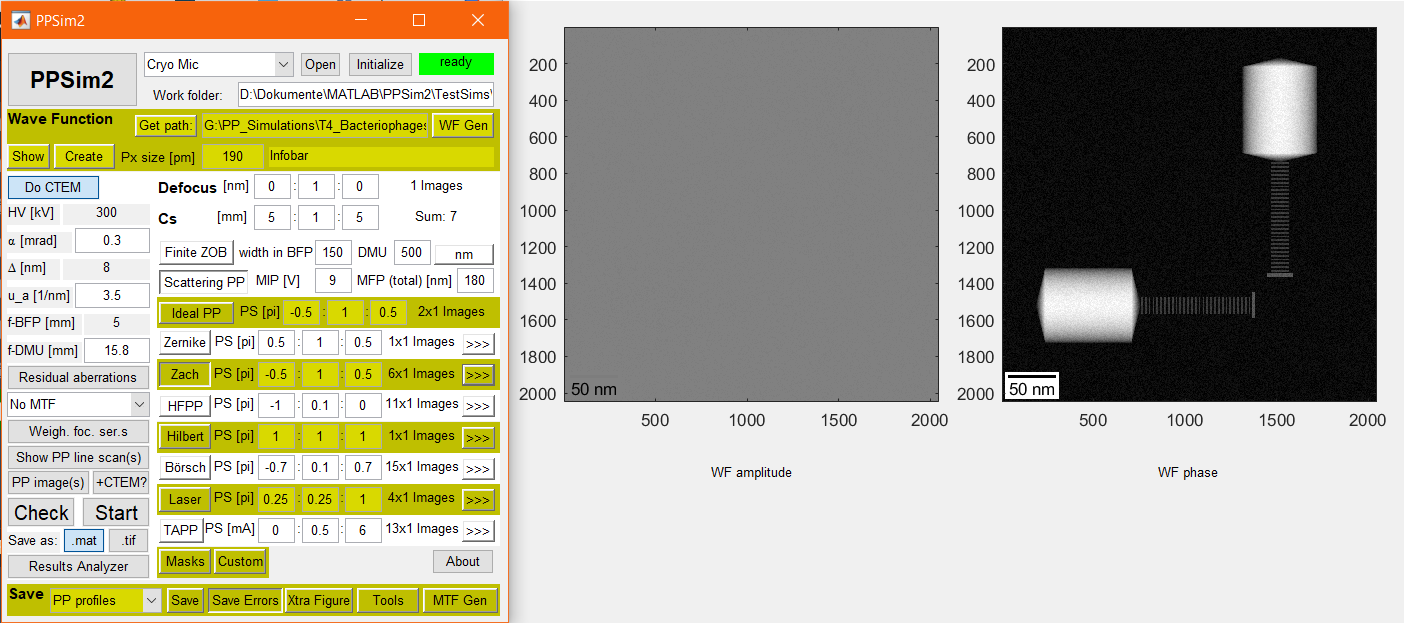}
    \caption{Screenshot of graphical user interface of the software used for the conducted simulations.}
    \label{SI:software}
\end{figure}

The used software \textit{PPSim2} has been published under open source GPL v3 and is available for free as source code for Matlab and as compiled version for Windows under https://www.hettlers.eu/matlab/. The compiled version requires the Matlab Runtime 2017b\footnote{The runtime can be freely downloaded under https://www.mathworks.com/products/compiler/matlab-runtime.html (checked in May, 2021).}. Figure~\ref{SI:software} shows an example screenshot of the main part of the graphical user interface (GUI). The program exhibits the following main features:

\begin{itemize}
    \item Simulation of conventional and PP TEM images based on linear image formation with optional consideration of partial coherence either by the weighted focal-series method or using simple damping envelopes. The following types of PPs can be considered:
    \begin{itemize}
        \item Ideal PP
        \item Zernike PP \citep{NagayamaDanev2001b}: conventional and with smoothed edge
        \item Zach PP \citep{PPZach_MM2010}
        \item HFPP: patch created by charge or contamination \citep{HFPP_2012,VPP_2014}
        \item Hilbert PP \citep{Nagayama_JBP_2002}
        \item Boersch PP \citep{BoerschOptimize.2007}: with one, two, or three supporting arms
        \item Laser PP \citep{Glaeser_LaserPP_2019}
        \item Tunable Ampere PP \citep{Marco_TunablePP_2018}
        \item Obstructing masks (Foucault and Tulip) \citep{Cullis.1975,Glaeser.2013ChargingTulip}
        \item Custom PP by loading images for phase and amplitude
    \end{itemize}
    \item Wave functions of amorphous spherical NPs of arbitrary material described by mean inner potential and mean-free path, either on a substrate thin film or embedded in vitrified ice, can be generated. Wave functions can be created from image files for phase and amplitude. Wave function simulated by STEMsim \citep{Rosenauer.2008STEMSim} may be directly loaded.
    \item The PP can be placed in virtual microscopes described by various parameters, which can possess a diffraction-magnifying unit leading to two back-focal planes with distinct focal lengths. Residual aberrations including up to fivefold astigmatism, axial coma and axial star can be set. 
    \item 
    Custom MTFs can be generated from a beam-stop image and several MTFs \citep{Ruskin.2013} are available upon request.
    \item Optional consideration of a finite ZOB size and scattering in opaque PPs is available.
    \item The software allows variation of multiple variables during the simulation process, including the defocus, spherical aberration constant, induced phase shift and the cut-on frequency (if applicable). Other important parameters such as the  semi-convergence angle, the ZOB size, scattering properties of thin-film PPs or misplacement of PPs can be varied automatically in a parameter sweep.
    \item PP profiles can be displayed and exported as images and the software comes with a basic PCTF plotting feature.
    \item Simulated images can be saved as .tif or as .mat (Matlab) files. The images can be visualized and contrast and visibility may be analyzed of rectangular or circular image areas. Different noise levels can be added to the simulated images.
    \item GUI settings are automatically saved with the simulated images and may conveniently be saved and loaded.
\end{itemize}

\bibliographystyle{citstyle.bst}
\bibliography{Biblio.bib}

\begin{thebibliography}{10}
\expandafter\ifx\csname url\endcsname\relax
  \def\url#1{\texttt{#1}}\fi
\expandafter\ifx\csname urlprefix\endcsname\relax\def\urlprefix{URL }\fi
\expandafter\ifx\csname href\endcsname\relax
  \def\href#1#2{#2} \def\path#1{#1}\fi

\bibitem{Hettler.2021}
S.~Hettler, R.~Arenal, Comparative image simulations for phase-plate
  transmission electron microscopy, Ultramicroscopy (2021) 113319\href
  {http://dx.doi.org/10.1016/j.ultramic.2021.113319}
  {\path{doi:10.1016/j.ultramic.2021.113319}}.

\bibitem{Boersch_PP_1947}
H.~Boersch, Über die {K}ontraste von {A}tomen im {E}lektronenmikroskop, Z.
  Naturforschung 2a (1947) 615--633.

\bibitem{PPReviewMalacHettler}
M.~Malac, S.~Hettler, M.~Hayashida, E.~Kano, R.~F. Egerton, M.~Beleggia, Phase
  plates in the transmission electron microscope: Operating principles and
  applications, Microscopy 70~(1) (2021) 75--115.
\newblock \href {http://dx.doi.org/10.1093/jmicro/dfaa070}
  {\path{doi:10.1093/jmicro/dfaa070}}.

\bibitem{HFPP_2012}
M.~Malac, M.~Beleggia, Kawasaki, M., P.~Li, R.~Egerton, Convenient contrast
  enhancement by a hole-free phase plate, Ultramicroscopy 118 (2012) 77--89.
\newblock \href {http://dx.doi.org/10.1016/j.ultramic.2012.02.004}
  {\path{doi:10.1016/j.ultramic.2012.02.004}}.

\bibitem{VPP_2014}
R.~Danev, B.~Buijsse, M.~Khoshouei, J.~Plitzko, W.~Baumeister, Volta potential
  phase plate for in-focus phase contrast transmission electron microscopy,
  PNAS 111~(44) (2014) 15635--15640.
\newblock \href {http://dx.doi.org/10.1073/pnas.1418377111}
  {\path{doi:10.1073/pnas.1418377111}}.

\bibitem{ZernikeLight}
F.~Zernike, Phase contrast, a new method for the microscopic observation of
  transparent objects, Physica 9~(7) (1942) 686--698.
\newblock \href {http://dx.doi.org/10.1016/S0031-8914(42)80035-X}
  {\path{doi:10.1016/S0031-8914(42)80035-X}}.

\bibitem{NagayamaDanev2001b}
R.~Danev, K.~Nagayama, Transmission electron microscopy with zernike phase
  plate, Ultramicroscopy 88~(4) (2001) 243--252.
\newblock \href {http://dx.doi.org/10.1016/S0304-3991(01)00088-2}
  {\path{doi:10.1016/S0304-3991(01)00088-2}}.

\bibitem{BoerschOptimize.2007}
E.~Majorovits, B.~Barton, K.~Schultheiß, F.~Pérez-Willard, D.~Gerthsen,
  R.~Schröder, Optimizing phase contrast in transmission electron microscopy
  with an electrostatic (boersch) phase plate, Ultramicroscopy 107~(2) (2007)
  213 -- 226.
\newblock \href {http://dx.doi.org/10.1016/j.ultramic.2006.07.006}
  {\path{doi:10.1016/j.ultramic.2006.07.006}}.

\bibitem{PPZach_MM2010}
K.~Schultheiss, J.~Zach, B.~Gamm, M.~Dries, N.~Frindt, R.~R. Schr{\"o}der,
  D.~Gerthsen, New electrostatic phase plate for phase-contrast transmission
  electron microscopy and its application for wave-function reconstruction,
  Microscopy and Microanalysis 16~(6) (2010) 785--794.
\newblock \href {http://dx.doi.org/10.1017/S1431927610093803}
  {\path{doi:10.1017/S1431927610093803}}.

\bibitem{Marco_TunablePP_2018}
A.~H. Tavabi, M.~Beleggia, V.~Migunov, A.~Savenko, O.~{\"O}ktem, R.~E.
  Dunin-Borkowski, G.~Pozzi, Tunable ampere phase plate for low dose imaging of
  biomolecular complexes, Scientific reports 8~(1) (2018) 5592.
\newblock \href {http://dx.doi.org/10.1038/s41598-018-23100-3}
  {\path{doi:10.1038/s41598-018-23100-3}}.

\bibitem{Glaeser_LaserPP_2019}
O.~Schwartz, J.~J. Axelrod, S.~L. Campbell, C.~Turnbaugh, R.~M. Glaeser,
  H.~M{\"u}ller, Laser phase plate for transmission electron microscopy, Nature
  methods 16~(10) (2019) 1016--1020.
\newblock \href {http://dx.doi.org/10.1038/s41592-019-0552-2}
  {\path{doi:10.1038/s41592-019-0552-2}}.

\bibitem{ZPP_Glaeser_Boersch_UM_2007}
R.~Cambie, K.~H. Downing, D.~Typke, R.~M. Glaeser, J.~Jin, Design of a
  microfabricated, two-electrode phase-contrast element suitable for electron
  microscopy, Ultramicroscopy 107~(4-5) (2007) 329--339.
\newblock \href {http://dx.doi.org/10.1016/j.ultramic.2006.09.001}
  {\path{doi:10.1016/j.ultramic.2006.09.001}}.

\bibitem{ZPP_Boersch_UM_2010}
D.~Alloyeau, W.~Hsieh, E.~Anderson, L.~Hilken, G.~Benner, X.~Meng, F.~Chen,
  C.~Kisielowski, Imaging of soft and hard materials using a boersch phase
  plate in a transmission electron microscope, Ultramicroscopy 110 (2010)
  563--570.
\newblock \href {http://dx.doi.org/10.1016/j.ultramic.2009.11.016}
  {\path{doi:10.1016/j.ultramic.2009.11.016}}.

\bibitem{Pretzsch.2019}
R.~Pretzsch, M.~Dries, S.~Hettler, M.~Spiecker, M.~Obermair, D.~Gerthsen,
  Investigation of hole-free phase plate performance in transmission electron
  microscopy under different operation conditions by experiments and
  simulations, Advanced Structural and Chemical Imaging 5~(1) (2019) 5.
\newblock \href {http://dx.doi.org/10.1186/s40679-019-0067-z}
  {\path{doi:10.1186/s40679-019-0067-z}}.

\bibitem{PPFukudaCell_2015}
Y.~Fukuda, U.~Laugks, V.~Lučić, W.~Baumeister, R.~Danev, Electron
  cryotomography of vitrified cells with a volta phase plate, Journal of
  Structural Biology 190~(2) (2015) 143 -- 154.
\newblock \href {http://dx.doi.org/10.1016/j.jsb.2015.03.004}
  {\path{doi:10.1016/j.jsb.2015.03.004}}.

\bibitem{MatthiasEbolaNature2018}
Y.~Sugita, Y.~Kawaoka, T.~Noda, M.~Wolf, Structure of the ebola virus
  nucleocapsid core by single particle cryo-electron microscopy, Microscopy and
  Microanalysis 22~(S3) (2016) 66--67.
\newblock \href {http://dx.doi.org/10.1017/S1431927616001185}
  {\path{doi:10.1017/S1431927616001185}}.

\bibitem{2019_Kotani_SkyrmionHFPP}
A.~Kotani, K.~Harada, M.~Malac, H.~Nakajima, K.~Kurushima, S.~Mori, Magnetic
  textures in a hexaferrite thin film and their response to magnetic fields
  revealed by phase microscopy, Japanese Journal of Applied Physics 58~(6)
  (2019) 065004.
\newblock \href {http://dx.doi.org/10.7567/1347-4065/ab1b87}
  {\path{doi:10.7567/1347-4065/ab1b87}}.

\bibitem{SimonZachPP_2016}
S.~Hettler, M.~Dries, J.~Zeelen, M.~Oster, R.~R. Schr{\"o}der, D.~Gerthsen,
  High-resolution transmission electron microscopy with an electrostatic zach
  phase plate, New Journal of Physics 18~(5) (2016) 053005.
\newblock \href {http://dx.doi.org/10.1088/1367-2630/18/5/053005}
  {\path{doi:10.1088/1367-2630/18/5/053005}}.

\bibitem{Gamm.2008EffectPP}
B.~Gamm, K.~Schulthei{\ss}, D.~Gerthsen, R.~Schr{\"o}der, Effect of a physical
  phase plate on contrast transfer in an aberration-corrected transmission
  electron microscope, Ultramicroscopy 108~(9) (2008) 878--884.
\newblock \href {http://dx.doi.org/10.1016/j.ultramic.2008.02.009}
  {\path{doi:10.1016/j.ultramic.2008.02.009}}.

\bibitem{Gamm.2010}
B.~Gamm, M.~Dries, K.~Schultheiss, H.~Blank, A.~Rosenauer, R.~Schr{\"o}der,
  D.~Gerthsen, Object wave reconstruction by phase-plate transmission electron
  microscopy, Ultramicroscopy 110~(7) (2010) 807--814.
\newblock \href {http://dx.doi.org/10.1016/j.ultramic.2010.02.006}
  {\path{doi:10.1016/j.ultramic.2010.02.006}}.

\bibitem{Dries.2014}
M.~Dries, S.~Hettler, B.~Gamm, E.~M{\"u}ller, W.~Send, K.~M{\"u}ller,
  A.~Rosenauer, D.~Gerthsen, A nanocrystalline hilbert phase-plate for
  phase-contrast transmission electron microscopy, Ultramicroscopy 139 (2014)
  29--37.
\newblock \href {http://dx.doi.org/10.1016/j.ultramic.2014.01.002}
  {\path{doi:10.1016/j.ultramic.2014.01.002}}.

\bibitem{Obermair2020}
M.~Obermair, S.~Hettler, C.~Hsieh, M.~Dries, M.~Marko, D.~Gerthsen, Analyzing
  contrast in cryo-transmission electron microscopy: Comparison of
  electrostatic zach phase plates and hole-free phase plates, Ultramicroscopy
  218 (2020) 113086.
\newblock \href {http://dx.doi.org/10.1016/j.ultramic.2020.113086}
  {\path{doi:10.1016/j.ultramic.2020.113086}}.

\bibitem{HaradaGrating_2020}
K.~Harada, M.~Malac, M.~Hayashida, K.~Niitsu, K.~Shimada, D.~Homeniuk,
  M.~Beleggia, Toward the quantitative the interpretation of hole-free phase
  plate images in a transmission electron microscope, Ultramicroscopy 209
  (2020) 112875.
\newblock \href {http://dx.doi.org/10.1016/j.ultramic.2019.112875}
  {\path{doi:10.1016/j.ultramic.2019.112875}}.

\bibitem{BeleggiaZernikeFormula2008}
M.~Beleggia, A formula for the image intensity of phase objects in zernike
  mode, Ultramicroscopy 108 (2008) 953--958.
\newblock \href {http://dx.doi.org/10.1016/j.ultramic.2008.03.003}
  {\path{doi:10.1016/j.ultramic.2008.03.003}}.

\bibitem{Fukuda.2009}
Y.~Fukuda, Y.~Fukazawa, R.~Danev, R.~Shigemoto, K.~Nagayama, Tuning of the
  zernike phase-plate for visualization of detailed ultrastructure in complex
  biological specimens, Journal of Structural Biology 168~(3) (2009) 476--484.
\newblock \href {http://dx.doi.org/10.1016/j.jsb.2009.08.011}
  {\path{doi:10.1016/j.jsb.2009.08.011}}.

\bibitem{ZPPOptimizing_NagayamaDanev_UM2011}
R.~Danev, K.~Nagayama, Optimizing the phase shift and the cut-on periodicity of
  phase plates for tem., Ultramicroscopy 111 (2011) 1305--1315.
\newblock \href {http://dx.doi.org/10.1016/j.ultramic.2011.04.004}
  {\path{doi:10.1016/j.ultramic.2011.04.004}}.

\bibitem{ZPPEdgcombe_2014}
C.~Edgcombe, Imaging of weak phase objects by a zernike phase plate,
  Ultramicroscopy 136 (2014) 154--159.
\newblock \href {http://dx.doi.org/10.1016/j.ultramic.2013.09.004}
  {\path{doi:10.1016/j.ultramic.2013.09.004}}.

\bibitem{PPSim2}
S.~Hettler, {PPSim2} - {A program for PP TEM} image simulation, accessed on
  21.05.2021, https://www.hettlers.eu/matlab/.

\bibitem{Rosenauer.2008STEMSim}
A.~Rosenauer, M.~Schowalter, {STEMSIM}---a new software tool for simulation of
  {STEM HAADF Z}-contrast imaging, in: A.~G. Cullis, P.~A. Midgley (Eds.),
  Microscopy of Semiconducting Materials 2007, Vol. 120 of Springer Proceedings
  in Physics, {Springer Netherlands}, Dordrecht, 2008, pp. 170--172.
\newblock \href {http://dx.doi.org/10.1007/978-1-4020-8615-1_36}
  {\path{doi:10.1007/978-1-4020-8615-1_36}}.

\bibitem{STEMCELLGrillo.2013}
V.~Grillo, E.~Rotunno, {STEM}{\_}{CELL}: A software tool for electron
  microscopy: part 1--simulations, Ultramicroscopy 125 (2013) 97--111.
\newblock \href {http://dx.doi.org/10.1016/j.ultramic.2012.10.016}
  {\path{doi:10.1016/j.ultramic.2012.10.016}}.

\bibitem{Barthel.2018}
J.~Barthel, Dr. {P}robe: A software for high-resolution stem image simulation,
  Ultramicroscopy 193 (2018) 1--11.
\newblock \href {http://dx.doi.org/10.1016/j.ultramic.2018.06.003}
  {\path{doi:10.1016/j.ultramic.2018.06.003}}.

\bibitem{harscher1998inelasticMIP}
A.~Harscher, H.~Lichte, Inelastic mean free path and mean inner potential of
  carbon foil and vitrified ice measured with electron holography, ICEM14,
  Cancun, Mexico 31 (1998) 553--554.

\bibitem{ReimerTEM_5}
H.~Kohl, L.~Reimer, Transmission electron microscopy: physics of image
  formation, Springer, 2008.

\bibitem{RosenauerBook.2003}
A.~Rosenauer, Transmission Electron Microscopy of Semiconductor Nanostructures:
  Analysis of Composition and Strain State, Vol. 182, {Springer Berlin
  Heidelberg}, Berlin, Heidelberg, 2003.
\newblock \href {http://dx.doi.org/10.1007/3-540-36407-2}
  {\path{doi:10.1007/3-540-36407-2}}.

\bibitem{Thust.2009}
A.~Thust, High-resolution transmission electron microscopy on an absolute
  contrast scale, Physical Review Letters 102~(22).
\newblock \href {http://dx.doi.org/10.1103/PhysRevLett.102.220801}
  {\path{doi:10.1103/PhysRevLett.102.220801}}.

\bibitem{HFPP_Pol2017}
M.~Malac, S.~Hettler, M.~Hayashida, M.~Kawasaki, Y.~Konyuba, Y.~Okura,
  H.~Iijima, I.~Ishikawa, M.~Beleggia, Computer simulations analysis for
  determining the polarity of charge generated by high energy electron
  irradiation of a thin film, Micron 100 (2017) 10--22.
\newblock \href {http://dx.doi.org/10.1016/j.micron.2017.03.015}
  {\path{doi:10.1016/j.micron.2017.03.015}}.

\bibitem{Egerton.2011}
R.~F. Egerton, \href{http://dx.doi.org/10.1007/978-1-4419-9583-4}{Electron
  Energy-Loss Spectroscopy in the Electron Microscope}, 3rd Edition, {Springer
  Science+Business Media LLC}, Boston, MA, 2011.
\newblock \href {http://dx.doi.org/10.1007/978-1-4419-9583-4}
  {\path{doi:10.1007/978-1-4419-9583-4}}.
\newline\urlprefix\url{http://dx.doi.org/10.1007/978-1-4419-9583-4}

\bibitem{Nagayama_JBP_2002}
R.~Danev, H.~Okawara, N.~Usuda, K.~Kametani, K.~Nagayama, A novel
  phase-contrast transmission electron microscopy producing high-contrast
  topographic images of weak objects, Journal of Biological Physics 28~(4)
  (2002) 627--635.
\newblock \href {http://dx.doi.org/10.1023/A:1021234621466}
  {\path{doi:10.1023/A:1021234621466}}.

\bibitem{2018_Simon_NegCharge}
S.~Hettler, E.~Kano, M.~Dries, D.~Gerthsen, L.~Pfaffmann, M.~Bruns,
  M.~Beleggia, M.~Malac, Charging of carbon thin films in scanning and
  phase-plate transmission electron microscopy, Ultramicroscopy 184 (2018)
  252--266.
\newblock \href {http://dx.doi.org/10.1016/j.ultramic.2017.09.009}
  {\path{doi:10.1016/j.ultramic.2017.09.009}}.

\bibitem{2019_Simon_LN2_Charging}
S.~Hettler, J.~Onodab, R.~Wolkow, J.~Pitters, M.~Malac, Charging of electron
  beam irradiated amorphous carbon thin films at liquid nitrogen temperature,
  Ultramicroscopy 196 (2019) 161--166.
\newblock \href {http://dx.doi.org/10.1016/j.ultramic.2018.10.010}
  {\path{doi:10.1016/j.ultramic.2018.10.010}}.

\bibitem{Cullis.1975}
A.~G. Cullis, D.~M. Maher, Topographical contrast in the transmission electron
  microscope, Ultramicroscopy 1~(2) (1975) 97--112.
\newblock \href {http://dx.doi.org/10.1016/S0304-3991(75)80012-X}
  {\path{doi:10.1016/S0304-3991(75)80012-X}}.

\bibitem{ZPP_Barton_Boersch_UM2011}
B.~Barton, D.~Rhinow, A.~Walter, R.~Schröder, G.~Benner, E.~Majorovits,
  M.~Matijevic, H.~Niebel, H.~Muller, M.~Haider, M.~Lacher, S.~Schmitz,
  P.~Holik, W.~Kuhlbrandt, In-focus electron microscopy of frozen-hydrated
  biological samples with a boersch phase plate, Ultramicroscopy 111~(12)
  (2011) 1696 -- 1705.
\newblock \href {http://dx.doi.org/10.1016/j.ultramic.2011.09.007}
  {\path{doi:10.1016/j.ultramic.2011.09.007}}.

\bibitem{Obermair2021GradedZPP}
M.~Obermair, Phasenkontrast-transmissionselektronenmikroskopie mit dünnfilm-
  und elektrostatischen phasenplatten, PhD thesis, Karlsruhe Institute of
  Technology\href {http://dx.doi.org/10.5445/IR/1000131823}
  {\path{doi:10.5445/IR/1000131823}}.

\bibitem{ZPP_cryo_tomo_Rado_2010}
R.~Danev, S.~Kanamaru, M.~Marko, K.~Nagayama, Zernike phase contrast
  cryo-electron tomography, Journal of Structural Biology 171~(2) (2010) 174 --
  181.
\newblock \href {http://dx.doi.org/DOI: 10.1016/j.jsb.2010.03.013}
  {\path{doi:DOI: 10.1016/j.jsb.2010.03.013}}.

\bibitem{BoerschPP_UM2015_Walter}
A.~Walter, S.~Steltenkamp, S.~Schmitz, P.~Holik, E.~Pakanavicius, R.~Sachser,
  M.~Huth, D.~Rhinow, W.~Kühlbrandt, Towards an optimum design for
  electrostatic phase plates, Ultramicroscopy 153 (2015) 22--31.
\newblock \href {http://dx.doi.org/10.1016/j.ultramic.2015.01.005}
  {\path{doi:10.1016/j.ultramic.2015.01.005}}.

\bibitem{ZPPOptimDesign2016}
D.~Rhinow, Towards an optimum design for thin film phase plates,
  Ultramicroscopy 160 (2016) 1 -- 6.
\newblock \href {http://dx.doi.org/10.1016/j.ultramic.2015.09.003}
  {\path{doi:10.1016/j.ultramic.2015.09.003}}.

\bibitem{Meyer.2007}
J.~C. Meyer, A.~K. Geim, M.~I. Katsnelson, K.~S. Novoselov, T.~J. Booth,
  S.~Roth, The structure of suspended graphene sheets, Nature 446~(7131) (2007)
  60--63.
\newblock \href {http://dx.doi.org/10.1038/nature05545}
  {\path{doi:10.1038/nature05545}}.

\bibitem{Butz.2014}
B.~Butz, C.~Dolle, F.~Niekiel, K.~Weber, D.~Waldmann, H.~B. Weber, B.~Meyer,
  E.~Spiecker, Dislocations in bilayer graphene, Nature 505~(7484) (2014)
  533--537.
\newblock \href {http://dx.doi.org/10.1038/nature12780}
  {\path{doi:10.1038/nature12780}}.

\bibitem{Ayala.2010}
P.~Ayala, R.~Arenal, M.~R{\"u}mmeli, A.~Rubio, T.~Pichler, The doping of carbon
  nanotubes with nitrogen and their potential applications, Carbon 48~(3)
  (2010) 575--586.
\newblock \href {http://dx.doi.org/10.1016/j.carbon.2009.10.009}
  {\path{doi:10.1016/j.carbon.2009.10.009}}.

\bibitem{Arenal.2014}
R.~Arenal, K.~March, C.~P. Ewels, X.~Rocquefelte, M.~Kociak, A.~Loiseau,
  O.~St{\'e}phan, Atomic configuration of nitrogen-doped single-walled carbon
  nanotubes, Nano letters 14~(10) (2014) 5509--5516.
\newblock \href {http://dx.doi.org/10.1021/nl501645g}
  {\path{doi:10.1021/nl501645g}}.

\bibitem{PPMikePractical_2016}
M.~Marko, C.~Hsieh, E.~Leith, D.~Mastronarde, S.~Motoki, Practical experience
  with hole-free phase plates for cryo electron microscopy, Microscopy and
  Microanalysis 22 (2016) 1316--–1328.
\newblock \href {http://dx.doi.org/10.1017/S143192761601196X}
  {\path{doi:10.1017/S143192761601196X}}.

\bibitem{Glaeser.2013ChargingTulip}
R.~M. Glaeser, S.~Sassolini, R.~Cambie, J.~Jin, S.~Cabrini, A.~K. Schmid,
  R.~Danev, B.~Buijsse, R.~Csencsits, K.~H. Downing, D.~M. Larson, D.~Typke,
  B.~Han, Minimizing electrostatic charging of an aperture used to produce
  in-focus phase contrast in the tem, Ultramicroscopy 135 (2013) 6--15.
\newblock \href {http://dx.doi.org/10.1016/j.ultramic.2013.05.023}
  {\path{doi:10.1016/j.ultramic.2013.05.023}}.

\bibitem{Ruskin.2013}
R.~S. Ruskin, Z.~Yu, N.~Grigorieff, Quantitative characterization of electron
  detectors for transmission electron microscopy, Journal of Structural Biology
  184~(3) (2013) 385--393.
\newblock \href {http://dx.doi.org/10.1016/j.jsb.2013.10.016}
  {\path{doi:10.1016/j.jsb.2013.10.016}}.

\end{thebibliography}

\end{document}